\documentclass[aps,prd,preprint,nofootinbib,nobibnotes]{revtex4}
\usepackage{graphicx}

\begin{document}

\title{ Gamma rays from the neutralino
dark matter annihilations in the Milky Way substructures }

\author{Xiao-Jun Bi}
\affiliation{Key laboratory of particle astrophysics, 
Institute of High Energy Physics, Chinese Academy of 
Sciences, P.O. Box 918-3, Beijing 100049, P. R. China}
\email{bixj@mail.ihep.ac.cn}

\begin{abstract}


High resolution simulations reveal that in the cold dark matter scenario
the structures form hierarchically and a large number of substructures
survive in the galactic halos. The substructures can be probed if they
emit gamma rays via dark matter annihilation.
We calculated the gamma ray fluxes from the dark matter annihilations
in the substructures of our Galaxy within the frame of the minimal
supersymmetric extension of the standard model. The uncertainties of
the prediction from both the low energy supersymmetry and especially
from the density profiles of dark matter in the substructures are carefully
investigated. The cumulative number of substructures emitting gamma rays above 
any given flux is calculated.  Detectability
of the gamma rays from the substructures is discussed. We propose the
viability to detect these signals through the ground 
large field of view detectors.

\end{abstract}


\maketitle

\section{Introduction}

Present observations\cite{pelmutter,wmap} strongly support 
a standard cosmology ($\Lambda$CDM) in which the Universe is 
spatially flat and its energy budget is balanced by  $\sim 4$\%
baryonic matter,  $\sim 23$\% cold, collisionless non-baryonic 
dark matter (CDM),
and $\sim 73$\% dark energy. In the $\Lambda$CDM cosmology, the luminous
galaxies and clusters of galaxies form within the halos of CDM where the dark
matter potential wells trap the baryonic gas, which eventually cools and
condense to form the galaxies. The DM halos are assumed to form
hierarchically bottom up via gravitational amplification of initial
fluctuations, generated during an early epoch of inflation with a nearly
scale invariant primordial power spectrum.
In this paradigm, small mass objects collapse first and merge into 
larger and larger halos over time. 

When a small halo merges into a large 
host system it is not immediately destroyed, but instead begins to 
orbit within the host, gradually losing mass to the parent halo
due to the action of the tidal force from the host, the dynamical 
friction and the close encounters with other subhalos. 
N-body simulation has been extensively used to study the 
merging history and the structure of the CDM halos.
As the recent development of fast algorithms to integrate the 
orbits of millions of particles, high resolution simulations
indicate that 
a fraction of about ~10\% of the total halo mass may have survived tidal
disruption and appear as distinct and self-bound substructures or
subhalos inside the virialized host halos\cite{tormen98,klypin99,moore99,
ghigna00,springel01,zentner03,delucia04,kravtsov04}.
The final configurations are self 
similar with a smooth host halo and a small fraction of the 
total mass in subhalos.
The halos that contain a wealth of substructures resemble the observed
clusters which host the galaxies, while the internal structure of a
galaxy-sized halo that hosts galaxy satellites looks just like a 
rescaled version of a rich cluster.

The existence of substructures in the halos 
have been confirmed by different high resolution numerical simulations
as a generic picture of the CDM cosmology with hierarchical structure
formation.  However, despite the great 
success of the CDM scenario in describing both the large scale
distribution of matter in our Universe\cite{wmap,sdss} and the structure of
galaxies and clusters\cite{klypin99,klypin03}, 
simulation shows that halos similar to that of 
Milky Way (MW) should
host hundreds of subhalos, which apparently overpredicts the 
abundance of substructures by an order of magnitude compared with
the 11 observed dwarf galactic satellites of the MW
\cite{kauffman93,klypin99,moore99}.
This CDM problem on sub-galactic scale is regarded as one of the 
most fundamental issues that has to be addressed. 

Several possible resolutions have been proposed to this apparent discrepancy.
One proposal is to change the nature of the dark matter, including
self-interacting dark matter model\cite{spergel00}, 
warm dark matter model \cite{colin00,bode01},
annihilating dark matter model \cite{kaplinghat00}, and
non-thermally produced dark matter model\cite{lin01,kamion04}.
Another possibility is to feature the inflation potential 
that suppress small scale power and thus reduce the predicted number of 
subgalactic halos\cite{Kamion2000,yokoyama00}.
However, these models become gradually disfavored by the recent
observations and numerical simulations \cite{hennawi02,yoshida03}.
The astrophysical mechanisms explain the discrepancy by suppressing 
dwarf galaxies formation in subgalactic halos\cite{AstroSol} and claim
that only very massive substructures contain stars and most
substructures are dark.
Therefore detection of the non-luminous subhalos in the galactic halos 
through lensing effects\cite{STRONG_LENSING}, tidal streams  
\cite{TIDAL_STREAM}, or any  other  method  
promise to distinguish between these alternatives. 
It seems that to account for the flux anomalies observed 
in radio lensing the amount of substructures predicted by the
$\Lambda$CDM model is required \cite{mao}. 

The subhalos may also be lit up by the annihilation of DM into
$\gamma$-rays and probed by $\gamma$-ray detectors if the DM 
particles are in form of weakly interacting particles \cite{Zeldovich}. 
The subhalos in MW greatly
enhance the fluxes of the annihilation products and
therefore enables us to detect these products,
since they produce many denser regions in the smooth background. 
To predict the intensity of the $\gamma$-ray fluxes it is necessary to
study the nature of the DM particles.

From the point of view of particle physics,
the existence of non-baryonic dark matter clearly
indicates the new physics beyond the standard model (SM) of particle physics.
Among the large amount of candidates proposed for non-baryonic DM
the leading scenario involves the weakly interacting massive particles 
(WIMPs), which is well motivated by theoretical extensions of the SM.
The weakly interacting relics from the early Universe with the WIMP mass 
from some tens of GeV to several TeV can naturally give rise to 
relic densities in the range of the observed DM density.
The most popular and natural extension of the SM seems to be
its supersymmetric (SUSY) version, i.e., the minimal supersymmetric
standard model (MSSM). The lightest supersymmetric particle (LSP) of the
MSSM, usually the neutralino, is neutral and stable due to R-parity 
conservation and provides an excellent candidate of CDM.
Search for dark matter via detection of its annihilation secondaries
is strongly motivated to unveil the form of new physics beyond the SM. 


Assuming that neutralino makes up the dark matter, in this paper 
we calculate the $\gamma$-ray fluxes from the dark matter 
annihilations in the MW subhalos. There have been similar studies
in the literature\cite{subworks,kou,Evans}. In this work 
we have addressed the uncertainties from
the unknown low energy SUSY parameters and from the not
well determined DM density profile in subhalos simultaneously. 
Especially we pay more attention on how to determine the distribution
and the density profile of the subhalos. 
Our results show that the prediction depends heavily on both sides of
particle physics and the cosmological evolution of the large scale structure.
The constraints on the DM annihilations from the observations of
EGRET\cite{egret}, CANGAROO II\cite{cang} and HESS \cite{hess} are taken
into account. 
The possibility of detection of these $\gamma$-rays is then discussed. 
Complementary to the space experiments, such as GLAST \cite{glast}
and the atmosphere
\v{C}erenkov detectors, such as VERITAS\cite{veritas}, MAGIC\cite{magic} or
HESS\cite{hessh},
we find the large ground cosmic ray arrays, such as the ARGO\cite{argo} and
the HAWC\cite{hawc} project may have the ability to detect the 
annihilations from the subhalos if the neutralino is as heavy as 
about 500 GeV and the density profile of subhalos has a steep
central cusp.
Therefore in most calculations we specify the quantities corresponding to
the ground array detectors, such as the angular resolution and 
the threshold energy.

The paper is organized as follows. In the Section II, we present the 
general formulas for calculating the $\gamma$-ray fluxes from DM 
annihilations. In the Section
III, we calculate the $\gamma$-ray flux from the DM annihilations 
at the Galactic Center (GC) and study the uncertainties from the `particle
factor' of Eq. (\ref{flux}). Since there are several observations at
the GC, we consider the constraints on the SUSY parameters from these
observations. 
In Section IV we present our results of the $\gamma$-ray fluxes from subhalos
after determining the distribution and density profile of the subhalos.
The detectability of the signals by different types of detectors is
discussed in Section V. In Section VI we give our conclusions.

\section{fluxes by dark matter annihilation}

It is easy to show that the average annihilation rate in unit time and 
unit volume is given by
\begin{equation}
R=\langle \sigma v \rangle n^2/2=\frac{\langle\sigma v\rangle\rho^2}{2m^2}
\end{equation}
where $\sigma$ and $v$ are the annihilation cross section and 
the relative velocity of the two dark matter particles respectively,
$n$ and $\rho$ are the number and mass densities of dark matter
and $m$ is its mass.
We note that the annihilation rate is proportional to the square
of the dark matter density and therefore,
a high density region can greatly enhances
the annihilation fluxes. 

The radiation fluxes from a dark matter halo is therefore given by
\begin{equation}
\label{flux}
\Phi(E)=\phi(E)\frac{\langle\sigma v\rangle}{2m^2}
\int{dV \frac{\rho^2}{4\pi d^2}} = \frac{\phi(E)}{4\pi}
\frac{\langle\sigma v\rangle}{2m^2}\times \int_{\Delta \Omega}d\Omega
\int_{\mathrm{l.o.s}}dl(r)\rho^2(r)
\end{equation}
in unit of $1$ particle $GeV^{-1} cm^{-2} s^{-1}$,
where $d$ is the distance from the detector to the source 
where dark matter annihilates
and $\phi(E)$ is the differential flux at energy $E$
by a single annihilation in unit of $1$ particle $GeV^{-1}$. 
At the last step of Eq. (\ref{flux}),
the integration is given alone the line-of-sight $l$, which is related
with the galactocentric distance $r$ 
by $r=\sqrt{l^2+r_\odot^2-2lr_\odot\cos\psi}$
with $r_\odot=8.5 kpc$ the distance of the Sun to the galactic center
and $\psi$ the direction of the source from the galactic center. 
$\Delta \Omega$ represents the solid angle at the direction of
$\psi$ for a given angular resolution of the the detector.
We notice that the integration in Eq. (\ref{flux}) depends
only on the distribution of the dark matter $\rho(r)$, taken as a 
spherically-averaged form,  which is determined
by numerical simulation or by observations and has no relation
to the particle nature of the dark matter. We define this factor as
`cosmological factor' and the other part in Eq. (\ref{flux}) 
the `particle factor' which is exclusively determined by its particle
nature, such as the mass, strength of interaction and so on. 
The factorization of the expression for
the annihilation fluxes into a cosmological part and a particle part
greatly simplifies our discussion. We will
discuss the two factors in the next sections one by one. 

\section{Gamma ray flux from the Galactic Center}

In this section we will study the uncertainties from the 
`particle factor' of Eq. (\ref{flux}) by calculating the 
$\gamma$-ray flux from dark matter
annihilations at the GC. This factor is exclusively determined
by particle physics and irrelevant to the source of the $\gamma$-rays.
Several relevant concepts will be introduced in this calculation. 
Since there are observations at the GC we will constrain
the SUSY parameter space from these observations. 

The Galactic Center is the most extensively studied region for
the dark matter annihilations \cite{works} and taken as the most
promising source to detect the annihilation products
since the density cusp at the GC can greatly enhance
the annihilation fluxes. However, due to the complexity of the
GC with the supermassive black hole and many baryonic
processes there is no convincing signal for the dark matter annihilation 
even though the excess of high energy $\gamma$-rays beyond the expected 
background have been detected by several experiments\cite{egret,cang,hess}.
On the contrary, if $\gamma$-rays are detected from the subhalos which
are otherwise completely dark it is a clear signal of the dark matter
annihilation. In the following sections we will see that there are other
advantages to detect dark matter annihilations from the MW substructures.

We first determine the `cosmological factor' of the GC in the subsection A
and then calculate the `particle factor' in the subsection B by scanning
the low energy SUSY parameter space and considering the constraints. Finally
we give our results in the subsection C.

\subsection{Cosmological factor}

\subsubsection{density profile of the MW}

The DM density profile is extensively studied by numerical simulations.
However, there are still a lot of debates on this subjects, which
focus on the slope of the central cusp of the profile. It is first 
given by Navarro, Frenk, and White \cite{nfw97} and supported by 
recent studies\cite{nfws} that the DM profile of isolated and relaxed halos
can be describe by a universal form 
\begin{equation}
\label{nfw}
\rho_{DM}(r)= \frac{\rho_s}{(r/r_s)(1+r/r_s)^2}\ ,
\end{equation}
where $\rho_s$ and $r_s$ are the scale density and scale radius respectively.
The two free parameters of the profile can be determined by 
the measurements of the virial mass of the halo 
and the concentration parameter determined by simulations.
The concentration parameter is defined as
\begin{equation}
c=\frac{r_{vir}}{r_{-2}}\ ,
\end{equation}
where $r_{vir}$ is  the virial radius of the halo and $r_{-2}$ is the radius 
at which the effective logarithmic slope of the
profile is $-2$, i.e., $\frac{d}{dr}(r^2\rho(r))\left|_{r=r_{-2}}=0\right . $.
For the NFW profile we have $r_s=r_{-2}$.
The concentration parameter reflects how the DM
is concentrated at the center. 
For a larger concentration parameter the DM is more centrally concentrated. 

The NFW profile in Eq. (\ref{nfw}) has a singularity when $r$ towards zero,
$\rho(r) \to r^{-1}$, while its slope becomes much steeper at large $r$,
$\rho(r) \to r^{-3}$ for $r \gg r_s$.  

However, Moore \textit{et al.} gave another form  of the DM 
profile \cite{moore} to fit their numerical simulation
with an increased resolution 
\begin{equation}
\label{moore}
\rho_{DM}(r)= \frac{\rho_s}{(r/r_s)^{1.5}(1+(r/r_s)^{1.5})}\ ,
\end{equation}
which has the same behavior at large radius as the NFW profile while 
it has a steeper central cusps $\rho(r) \to r^{-1.5}$ for small $r$ than
the NFW profile.
The index of the central cusp at about $1.5$ is also favored by following 
higher resolution simulations\cite{moores}.
At the present time it seems that the different central cuspy behavior
are not due to finite numerical resolution of the
simulations and can not be solved by improving the numerical 
resolution further. For the Moore profile we have $r_s=r_{-2}/0.63$.

Anyway, although there is on going debate over which profile is most accurate
it is reasonable to believe that the NFW and the Moore profiles
represent two limiting cases between which a realistic description
of DM distribution will fall. We will calculate the gamma ray fluxes
from neutralino annihilation by adopting the two profiles. The actual
fluxes should fall within the two limiting cases.

\subsubsection{core radius}

We notice that both the NFW and the Moore profiles have unphysical
singularities at the GC which may lead to divergent gamma ray fluxes. 
Therefore the profiles should have a core radius, $r_{\text{core}}$,
within which the DM profile should be kept constant due to the balance 
between the very high annihilation rate and the rate to fill the region
by infalling DM particles. 
The time scale of the
free fall of the DM particles can be approximately given by\cite{core}
\begin{equation}
\tau \sim \frac{1}{\sqrt{G\bar{\rho}}}\ ,
\end{equation}
while the annihilation time scale is 
\begin{equation}
\tau \sim \frac{1}{\langle \sigma v\rangle n_\chi(r_{\text{core}}) } \ .
\end{equation}
Taking $\bar{\rho}$ about 200 times the critical density and 
$\langle \sigma v\rangle\sim 10^{-26} cm^3 s^{-1}$ and applying 
the formulas above we then get
$r_{\text{core}}\sim 10^{-8}$ kpc for the Moore profile and
$r_{\text{core}}\sim 10^{-11}$ kpc for the NFW profile. 

\begin{figure}
\includegraphics[scale=0.5]{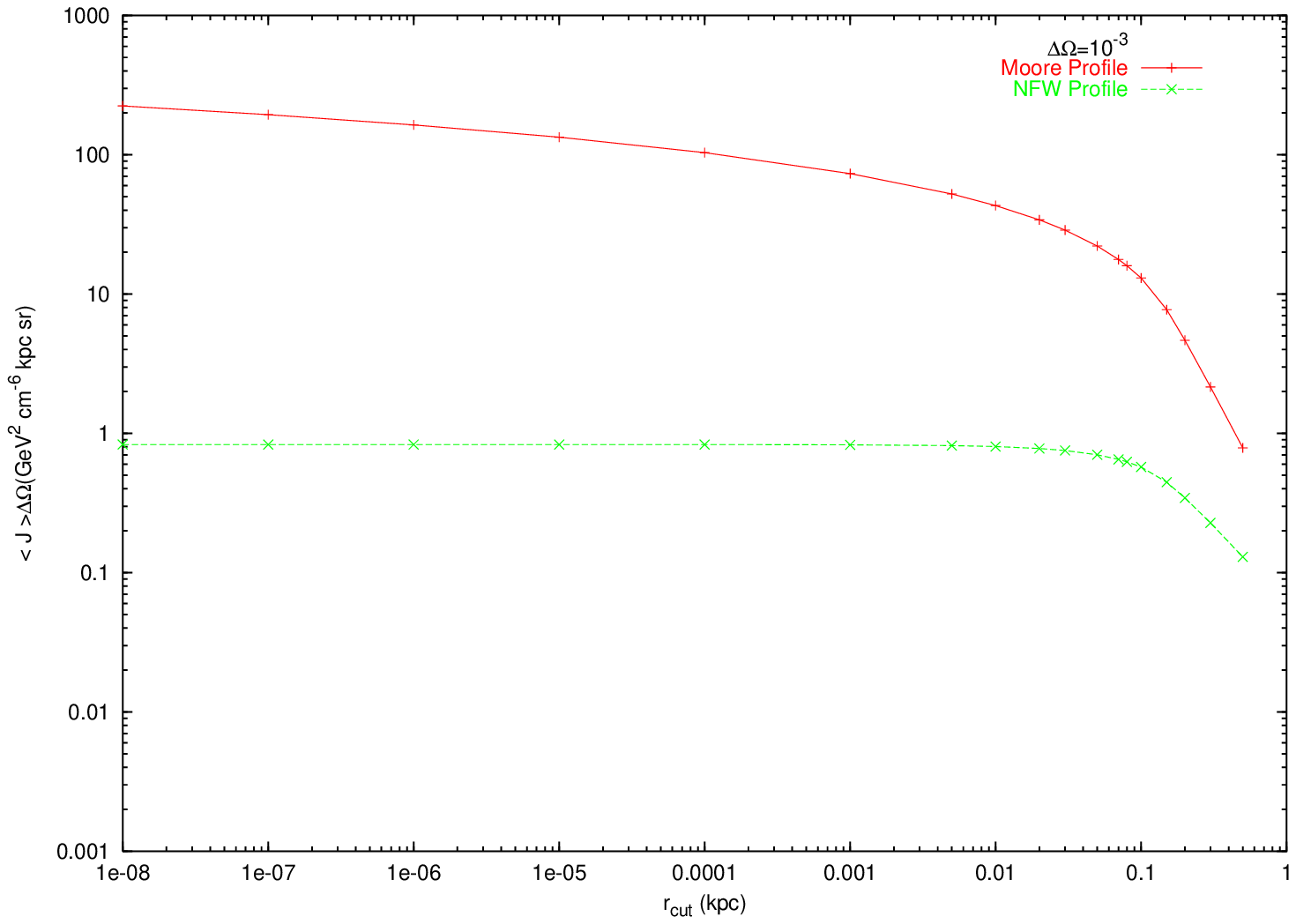}
\includegraphics[scale=0.5]{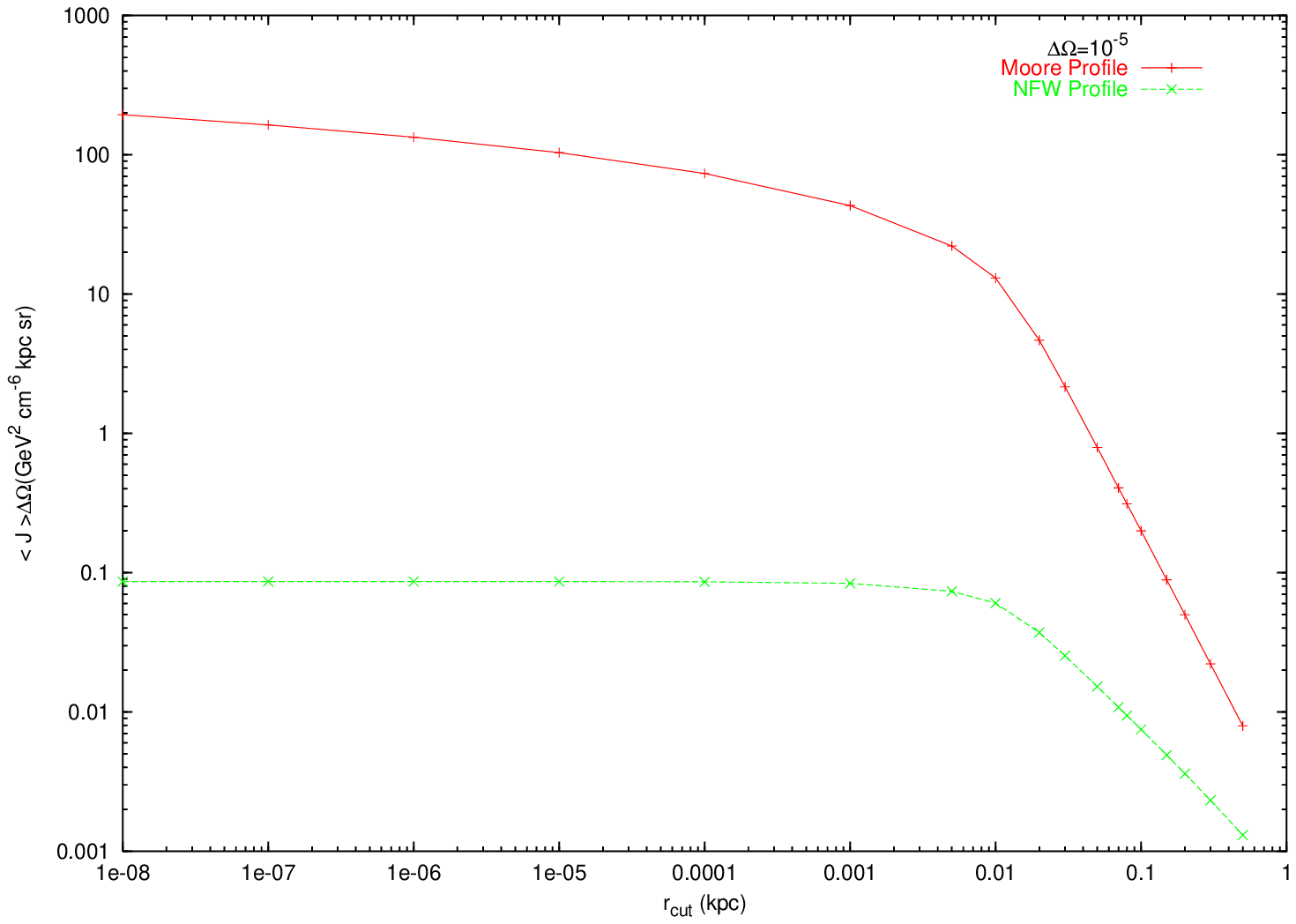}
\caption{\label{core_radius}
The cosmological factor within the solid angle 
$\Delta\Omega$ as function of the core radius
for the angular resolution of $\Delta\Omega=10^{-3},\ 10^{-5}$ respectively.
}
\end{figure}

At the position of the Sun, which is about 8.5 kpc away from the the 
Galactic Center, we can probe the GC 
to a radius as small as 0.15 kpc 
(0.015 kpc) using an instrument with the angular resolution of $1^\circ$ 
($0.1^\circ$), corresponding to the solid angle
$\Delta\Omega\approx 10^{-3}(10^{-5})$. 
Therefore we expect the annihilation flux within the solid angle 
$\Delta\Omega$ will
be enhanced greatly if the core radius is smaller than the radius
that the instrument can resolve, i.e., $r_{\text{core}} < r_{\text{res}}$. 
In Fig. \ref{core_radius},
we plot the cosmological factor defined as the integration 
in Eq. (\ref{flux}) 
as a function of the core radius. From the figure we can see
that the cosmological factor increases quickly with decreasing
core radius when it is larger than the $r_{\text{res}}$. For smaller
$r_{\text{core}} < r_{\text{res}}$, the cosmological factor rises
slowly by further decreasing the core radius. This behavior can be
understood as below. The flux is proportional to integration of
the density square within the solid angle 
\begin{equation}
\Phi \sim \int_0^{r_{\text{res}}} \rho^2(r) r^2dr\sim
 \int_0^{r_{\text{core}}} \rho^2(r_{\text{core}}) r^2dr
+\int_{r_{\text{core}}}^{r_{\text{res}}} \rho^2(r)r^2dr\ ,
\end{equation}
which gives that the flux in Moore profile depends on $r_{\text{core}}$
logarithmically while independent of $r_{\text{core}}$ for the NFW case.

\subsubsection{angular resolution}

From Eq. (\ref{flux}) we can see that the cosmological factor
also depends on the angular resolution of the experimental instruments.
We expect a larger cosmological factor for a larger angular resolution
which probes greater volume of dark matter annihilation (however, the
significance of detection is decreased due to more background is
included. see our later discussion). In Fig. 
\ref{core_radius}, we plot the cosmological factor for the solid 
angular resolution $\Delta\Omega = 10^{-3}$ and $\Delta\Omega = 10^{-5}$
respectively. For the case when $r_{\text{core}} > r_{\text{res}}$
the cosmological factor for $\Delta\Omega = 10^{-3}$ is indeed
two orders magnitude larger than that for $\Delta\Omega = 10^{-5}$.
While for smaller $r_{\text{core}}$ the difference between the two cases
is smaller. Especially for the Moore profile the difference is almost 
negligible.

\subsection{Particle factor}

We now turn to the particle factor in Eq. (\ref{flux}).
We will mainly work in the frame of MSSM, which is a low energy 
effective description of the fundamental theory at the
electroweak scale. 
For comparison, the minimal supergravity (mSUGRA) model is also explored,
where the soft SUSY breaking parameters can be universally defined at
the scale of grand unification (GUT).
For the R-parity conservative MSSM, 
the lightest supersymmetric particle (LSP), generally the 
lightest neutralino, is stable and is an ideal candidate of dark matter. 

Even the  R-parity conservative MSSM
is described by more than one hundred soft supersymmetry breaking 
parameters. However, for
the processes related with dark matter production and annihilation,
only several parameters are relevant under some simplifying assumptions,
i.e., the higgsino mass parameter $\mu$, the wino mass parameter $M_2$,
the mass of the CP-odd Higgs boson $m_A$, the ratio of the Higgs
Vacuum expectation values $\tan\beta$, the scalar quark mass parameter
$m_{\tilde{q}}$, the scalar lepton mass parameter $m_{\tilde{l}}$, 
the trilinear soft breaking parameter $A_t$
and $A_b$. To determine the low energy spectrum of the SUSY particles
and coupling vertices, 
the following assumptions have been made: all the sleptons
and the squarks have common soft-breaking mass parameters $m_{\tilde{l}}$
and $m_{\tilde{q}}$ respectively; all trilinear 
parameters are zero except those of the third family; the bino and wino
have the 
mass relation, $M_1=5/3\tan^2\theta_W M_2$, coming from the unification 
of the gaugino mass at the grand unification scale.

We perform a numerical random scanning of the 8-dimensional 
supersymmetric parameter space using the package DarkSUSY 
\cite{darksusy}. The ranges of the parameters are as following:
$50 GeV < |\mu|,\ M_2,\ M_A,\ m_{\tilde{q}},\ m_{\tilde{l}} < 5 TeV$, 
$1.1 < \tan\beta < 55$, $-3m_{\tilde{q}} < A_t, \ A_b < 3m_{\tilde{q}}$, 
$\text{sign}(\mu)=\pm 1$. 
The parameter space is constrained by the theoretical consistency 
requirement, such as the correct vacuum breaking pattern, 
the neutralino
being the LSP and so on. The accelerator data 
constrains the parameter further
from the spectrum requirement, the invisible Z-boson width, 
the branching ratio
of $b\to s\gamma$ and so on adopting the 2002 limits of the Particle
Data Group\cite{pdg}.

Another important constraint comes from cosmology. Combining the recent
observation data on cosmic microwave background, large scale structure,
supernova and data from HST Key Project 
the cosmological parameters are determined quite precisely. Especially,
the abundance of the cold dark matter is given by \cite{wmap} 
$\Omega_{\text{CDM}}h^2=0.113^{+0.008}_{-0.009}$. 
We constrain the SUSY parameter
space by requiring the relic abundance of neutralino 
$0 < \Omega_\chi h^2< 0.137$,
where the upper limit corresponds to the 3$\sigma$ upper bound from
the cosmological observations. When the relic abundance of neutralino
is smaller than a minimal value the neutralino represents a subdominant
dark matter component. We then rescale the galaxy dark matter density
as $\rho(r)\to\xi\rho(r)$ with 
$\xi=\Omega_\chi h^2/(\Omega_\chi h^2)_{\text{min}}$. We take 
$(\Omega_\chi h^2)_{\text{min}}=0.086$, the 3$\sigma$ lower bound
of the CDM abundance \cite{wmap}.
The effect of coannihilation between the fermions is taken into account
when calculating the relic density numerically. 

Besides exploring the SUSY parameter space at low energy directly
under some simplifying assumptions,
there is another popular approach of handling the phenomenology of MSSM by
assuming a simple SUSY breaking pattern at the GUT scale.
One of the simplest models in this kind is the minimal supergravity 
model which has only five
free SUSY parameters, the gaugino masses $m_{1/2}$, 
the sfermion masses $m_0$, the trilinear parameter $A_0$, which are all
defined universally at the GUT scale,  as well as the ratio of the
vacuum expectation values $\tan\beta$ and
the sign of the higgino mass parameter $\mu$.
The high energy universal relationship leads to constraints
on the low energy spectrum \cite{ellis}.
To calculate the neutralino annihilation, we adopt the 
package ISASUGRA (version 7.69) to calculate the low energy SUSY
spectrum by numerical solving the renormalization group 
equations from the GUT scale downwards to the weak scale \cite{isajet}. 
We also randomly scan the free parameter
space considering the theoretical and experimental constraints.
The ranges of the parameters are given as $50 GeV < m_0, m_{1/2} < 5 TeV$,
$-3m_0 < A_0 < 3m_0$, $1.1 <\tan\beta <55$ and $\text{sign}(\mu)=\pm 1$.

The  $\gamma$-rays from the neutralino annihilation arise mainly
in the decay of the neutral pions produced in the fragmentation processes
initiated by tree level final states. The fragmentation and decay processes 
are simulated with Pythia package\cite{pythia} 
incorporated in DarkSUSY. We focus our calculation 
on the continuum $\gamma$-rays from the pion decays.

\subsubsection{constraints from cosmic ray observations}
 
Before presenting the numerical results we discuss the 
constraints from the cosmic ray observations first. 
High energy and very high energy $\gamma$-ray emission 
from the GC have been detected by the EGRET \cite{egret},
CANGAROO-II \cite{cang} and HESS \cite{hess} experiments.
However, the present situation is still confused since all these results
can not be attributed to a unique $\gamma$-ray source or due to a single
emission mechanism. Therefore we take all these results as
a constraint on the dark matter annihilation, i.e., 
the predicted $\gamma$-rays flux due to the DM annihilation
should not exceed these observed fluxes.

The $\gamma$-ray spectrum from the EGRET observation\cite{egret} 
at the GC with the angular resolution of $\sim 1^\circ$ is well
described by a broken power law with a break energy at $1.9 GeV$. Above 
this energy the photon spectrum is 
$F(E)=1.6\times 10^{-6} (E/GeV)^{-3.1} cm^{-2} s^{-1} GeV^{-1}$. 
Assuming that this spectrum can extend to a quite high energy we get
an approximate integrated $\gamma$-ray flux above 
$1 GeV$ using this power spectrum
$F(E > 1 GeV) =7.7\times 10^{-7} cm^{-2}s^{-1}$.

Both CANGAROO-II and HESS observed $\gamma$-rays at the GC from
about $200 GeV$ to about $10 TeV$. The HESS data \cite{hess} is
fitted by a power law spectrum $F(E)=2.5\times 10^{-12} (E/TeV)^{-2.21}
cm^{-2}s^{-1}TeV^{-1}$. The CANGAROO-II \cite{cang,Hooper} gives a quite different
power law spectrum with a spectral index $-4.6$.
The apparent discrepancy seems indicate a significant change of the
source at lower energies
over about one year. However, none of the individual
experiment observes significant variability of the source. Another possible
explanation is that due to larger positional uncertainty
of the CANGAROO-II there may be more than one source exist in
the direction of the GC\cite{horns}. The flux detected by HESS is also
much lower than that detected by EGRET if extending the spectrum 
to lower energies. Therefore we take a
relaxed constraint from the HESS experiment, which has
the best angular resolution of $0.1^\circ$: assuming that the
source at the GC can extend to the range within an angular resolution
of $1^\circ$,
despite a point-like source can fit the single HESS observation well\cite{hess}.
We then get the integrated flux above $100 GeV$ using the given spectrum
$F(E > 100 GeV)=3.4\times 10^{-9} cm^{-2}s^{-1}$.
The integrated flux from the
CANGAROO-II data by extending its spectrum to lower energy,
$F(E > 100 GeV)=2.8\times 10^{-9} cm^{-2}s^{-1}$.
Anyway, at the moment the situation is not clear we take these values
as a conservative constraints so that we can probe more SUSY parameter
space. For further more severe constraints more SUSY parameter space
will be constrained and our later results can be
simply rescaled due to Eq. (\ref{flux}).

\subsection{Results}

\begin{figure}
\includegraphics[scale=0.7]{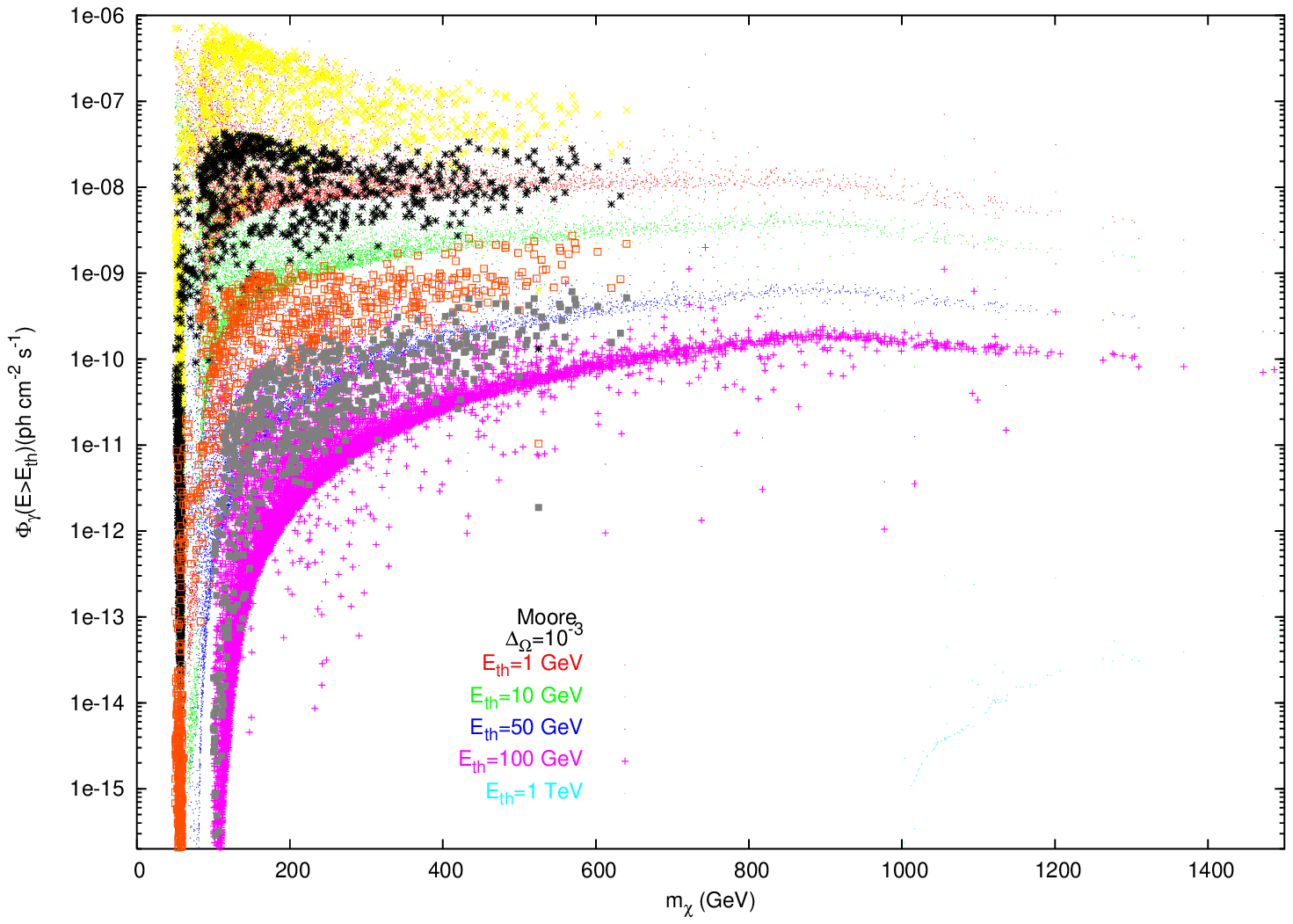}
\includegraphics[scale=0.7]{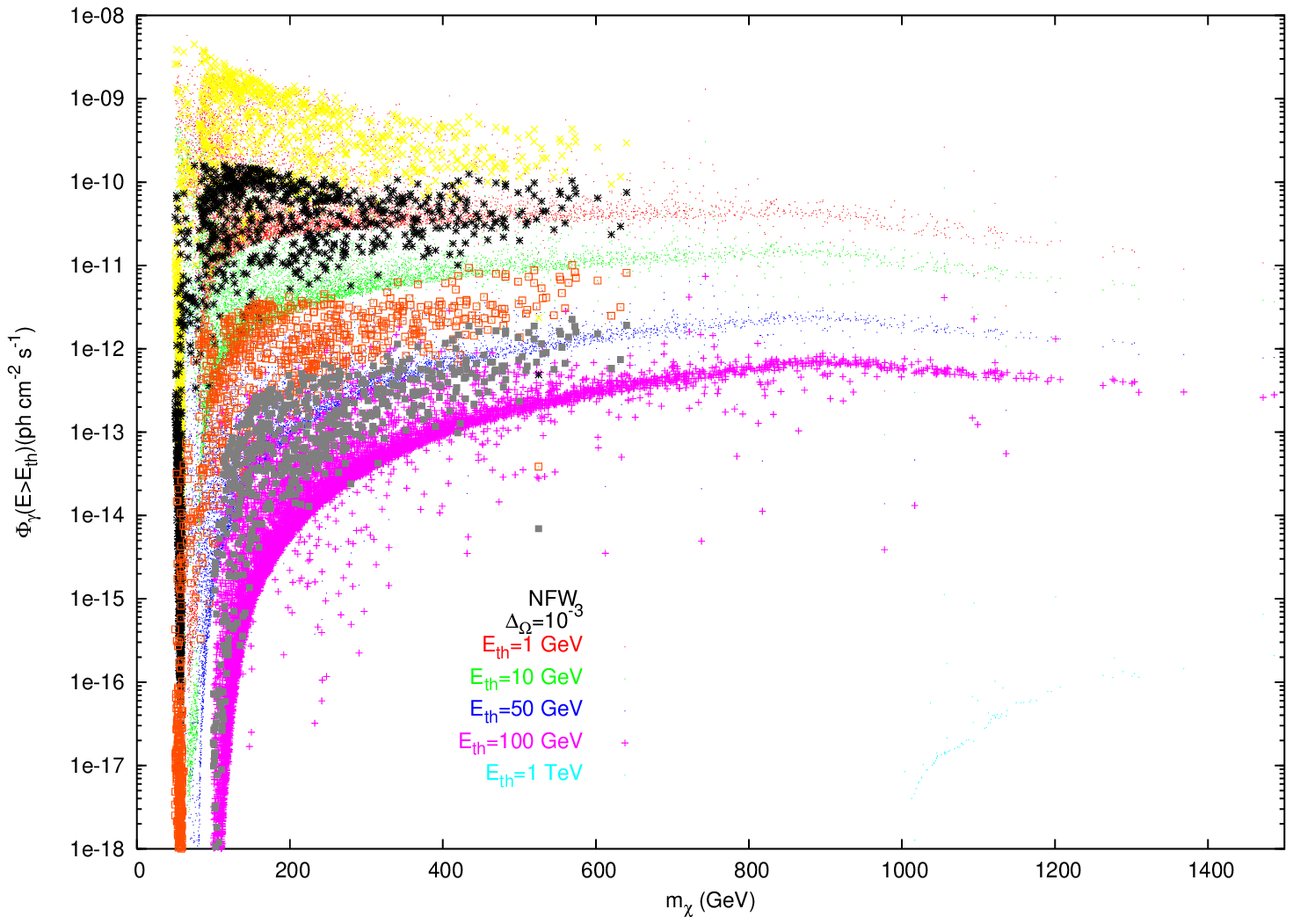}
\caption{\label{GC_flux}
The integrated $\gamma$-ray fluxes from neutralino annihilation at the
GC above the threshold energy of
$1$, $10$, $50$, $100$ and $1000$ GeV respectively 
for the Moore (upper panel) 
and the NFW (lower panel) profiles as function of the neutralino mass. 
The fluxes are given within the angular resolution of 
$\Delta\Omega =10^{-3}$. Each dot in the figure represents a set
of low energy SUSY parameters which survive all the current limits. 
The bigger points superposed on each set of dots (corresponding to
a threshold energy) are given by the
mSUGRA model.
}
\end{figure}

Combining all these constraints together, 
from the theoretical consideration to the
accelerator experiments and cosmic ray observations, we can give the
predicted $\gamma$-ray fluxes from neutralino
annihilations at the GC now. 
In Fig. \ref{GC_flux}, we plot the integrated 
$\gamma$-ray fluxes  within the solid angle
$\Delta\Omega=10^{-3}$ as a function of the neutralino mass taking
the threshold energy as $1 GeV$, $10 GeV$, $50 GeV$, $100 GeV$ and
$1TeV$ for both the Moore and the NFW profiles.
Each dot in the figure corresponds to a 
model with a set of definite SUSY parameters
in the 8-dimensional parameter space 
allowed by all the constraints. 
The bigger points superposed on each set of dots are given
by the mSUGRA model. We notice that the mSUGRA model tends
to give large annihilation fluxes, while small neutralino mass
which is hard to be greater than $800 GeV$. 
The scatter of the
points represents the uncertainty coming from the unknown 
soft SUSY breaking parameters.

It is worth giving some comments here. First, the core radius is taken
as $10^{-8}$ kpc for both the Moore and the NFW profiles. From
Fig. \ref{core_radius} we can see that the flux in the Moore profile
decrease only by a factor of 2 when taking 
$r_{\text{core}}$ from $10^{-8}$ kpc to $10^{-4}$ kpc. For the case of
NFW profile the predicted flux has no change even taking
$r_{\text{core}}$ as large as $0.1$ kpc.
Second, in the case of Moore profile the SUSY parameter space 
has been constrained by the cosmic ray observations at the GC. 
If more stringent constraints are adopted all the later results
according to the Moore profile will be rescaled.
However, there is no constraints for the case of the NFW profile.
Third, except the region near the threshold energy
the uncertainties of the predicted 
$\gamma$-ray fluxes are within two orders of magnitude
even scanning the 8-dimensional parameter space
which spans quite a large range, as given in the last subsection. 
The most strong constraint comes
from the requirement by cosmology, i.e., requiring 
$0 < \Omega_{\text{CDM}}h^2 < 0.137$. For s-wave annihilation
the rate
$\langle \sigma v\rangle_{\text{ann}}$ is closely related with the
initial value of $\langle \sigma v\rangle_{\text{ini}}$ at the
decoupling epoch which determines the relic density of neutralino.
Forth, the neutralino mass above $1\ TeV$ is difficult to achieve
after taking all the constraints into account. 
We can realize a model with the neutralino mass at most as heavy as 
about $5\ TeV$ to satisfy all the constraints after we relax the range
of all low energy soft mass parameters as large as $50 TeV$. 
Therefore, a neutralino
as heavy as $18\ TeV$ to explain the spectrum observed by HESS experiment
\cite{horns,profumo,Ferrer} as neutralino annihilation will
be very hard to achieve in the framework of MSSM.

For discussions of the substructure emission in the next sections
we will fix the `particle factor' by taking an optimistic set of SUSY
parameters.
The `particle factor' is such taken that the integrated $\gamma$-ray
fluxes above $100 GeV$ from the GC for the Moore profile is  
$F(E>100 GeV)=10^{-9} cm^{-2}s^{-1}$, which is near the maximal value
of the SUSY prediction. 
This is just a normalization of the $\gamma$-ray flux and only 
the relative magnitude of the fluxes 
between the GC and the substructure is relevant, since
the overall magnitude can be rescaled according to different particle
factors.

We can summarize this section here. By scanning the SUSY parameter space
and after taking all the constraints into account we give the scatter of 
the integrated $\gamma$-ray fluxes from neutralino annihilations 
at the GC in Fig. \ref{GC_flux} for different threshold energies and for
both the Moore and the NFW profiles. 
From the factorization of the expression in Eq. (\ref{flux}) we know
this figure is suitable to any other source by multiplying each point
by a global factor which represents the difference of the `cosmological factor'
from that at the GC. Therefore the figure shows the general uncertainties 
of the particle factor. 
The uncertainty is quite small considering the huge volume of the free
parameter space. This is 
due to the fact that the annihilation process is closely related with
the process of dark matter freezing out which determines its relic density.

\section{Gamma rays from the subhalos}

\subsection{realization of MW with substructures}

To predict the $\gamma$-ray flux by
neutralino annihilation from the subhalos we need to
know the distribution of subhalos in the MW
and the density profile within each subhalo.

The properties of subhalos are determined via the competition between
accretion and destruction due to tidal force and dynamical friction.
N-body simulation and semi-analytical methods have been extensively
used to investigate the spatial distribution and mass function of
substructures in the host halo.
According to the extensive studies it is now believed that the radial 
distribution of substructures is generally 
shallower than density profile of the smooth background.
The reason for the anti-bias of the substructure
number density relative to the smooth distribution is due to
the tidal disruption of substructures
which is most effective near the galactic center.
This conclusion does not depend on
the numerical resolution as confirmed in Ref. \cite{diemand} by adopting a
wide range of mass and force resolutions. 
It is shown that the relative number
density of subhalos can be approximated by an isothermal
profile with a core \cite{diemand}
\begin{equation}
\label{dis}
n(r)=2n_H(1+(r/r_H)^2)^{-1}\ ,
\end{equation}
where $n_H$ is the relative number density at the scale
radius $r_H$. The average core radius for the distribution
of galaxy subhalos is about $0.14$ times the halo virial radius, 
$r_H=0.14r_{\text{vir}}$. The core radius is a smaller fraction of the
virial radius than that of cluster subhalos\cite{diemand}, since galaxy
forms earlier and is more centrally concentrated. The spatial distribution
given above agrees well with that in another recent
simulation by Gao et al. \cite{gao}.

At large radius $n(r)$ in Eq. (\ref{dis}) goes
as $r^{-2}$ which represents the anti-bias since the DM profile goes
as $r^{-3}$ at large radius for both the NFW and the Moore profiles. 
It is worth mentioning that most previous works
calculating $\gamma$-rays from substructures adopting a fitted formula 
given in Ref. \cite{blasi}, $n(r)\propto (1+(r/r_H)^2)^{-1.5}$, which
follows the DM profile and does not reflect 
the anti-bias of the substructure distribution.

Simulations show that the differential mass 
function of substructures has an approximate power law 
distribution, $dn/dm\sim m^{-\alpha}$, 
with no dependence or slight dependence on the host
halo mass. Most studies point out that  the cluster and galaxy 
have similar substructure mass function although the 
clusters form much later than galaxies in the hierarchical
structure formation scenario. 
It seems that the tidal effects change the mass
distribution function self-similarly. 
In Ref. \cite{diemand} both the cluster and galaxy substructure
cumulative mass functions are found to be an $m^{-1}$ power law,
$n_{\text{sub}}(m_{\text{sub}} > m)\propto m^{-1}$,
with no dependence on the mass of the parent halo. 
A slight difference is found in a recent simulation by Gao \textit{et al.}
\cite{gao} that the cluster substructure is more abundant than galaxy 
substructure since the cluster forms later and more substructures 
have survived the tidal disruption.
The mass function for both scales are well fitted by $dn/dm\propto m^{-1.9}$.
A non-universal form of the mass function is found in Ref. \cite{ghigna00}:
the power index of the cumulative mass function is $-1$ for 
$m_{\text{sub}} > 10^{11}h^{-1}M_{\odot}$
while it changes to $-0.7$ for $m_{\text{sub}} < 10^{11}h^{-1}M_{\odot}$.
There is an advantage of a power law form for the differential mass function
shallower than $m^{-2}$: the fraction of the total mass enclosed in
subhalos is then insensitive to the mass of the minimal subhalo we take.
The mass fraction of subhalos estimated in the literature is around 
between 5 percent to 20 percent \cite{ghigna00,springel01,stoehr}.
In this work we will always take the mass fraction of substructures as 
10 percent.

Putting all these arguments together, we get the probability of a
substructure with mass $m$ at the position $r$ to the galactic center
\begin{equation}
\label{prob}
n(m,r)=n_0 \left(\frac{m}{M_{\text{vir}}}\right)^{-1.9} (1+(r/r_H)^2)^{-1}\ ,
\end{equation}
where $M_{\text{vir}}$ is the virial mass of the MW, $n_0$ is the 
normalization factor determined by requiring the total mass of 
substructures converges to 10 percent of the MW virial mass, $M_{\text{vir}}$. 
A population of substructures within the virial radius of the 
MW are then realized statistically due to Eq. (\ref{prob}). 
The mass of the substructures are taken randomly between 
$M_{\text{min}}=10^6 M_\odot$, which is the lowest substructure
mass the present simulations can resolve \cite{cut}, and the maximal mass 
$M_{\text{max}}$. 
The maximal mass of substructures is taken to be $0.01M_{\text{vir}}$ 
since the MW halo does not show recent mergers of satellites with masses
larger then $\sim 2\times 10^{10}M_\odot$.
It will be shown that the $\gamma$-ray flux is quite insensitive to 
the minimum subhalo mass since the flux from a single subhalo
scales as its mass \cite{kou,aloisio}.

We notice that the number density of substructures is largest at the GC due to  
Eq. (\ref{prob}) which, however, is in conflict with the fact that 
most substructures near the GC are destroyed by the strong tidal effect.
The underestimate of the tidal effect near the GC in 
Eq. (\ref{prob}) is due to the finite resolution 
of the N-body simulations and the formula is an extrapolation of the
subhalo distribution to smaller radius.
The global tides from the host halos strip outer parts of the
substructures and result in the substructure disruption or at least
significant amount of substructure mass loss. 
We take the tidal effects into
account under the ``tidal approximation'', which assumes that 
all mass beyond a suitably defined tidal radius is lost in a single
orbit while keep its density profile inside the tidal radius intact.

\begin{figure}
\includegraphics[scale=0.7]{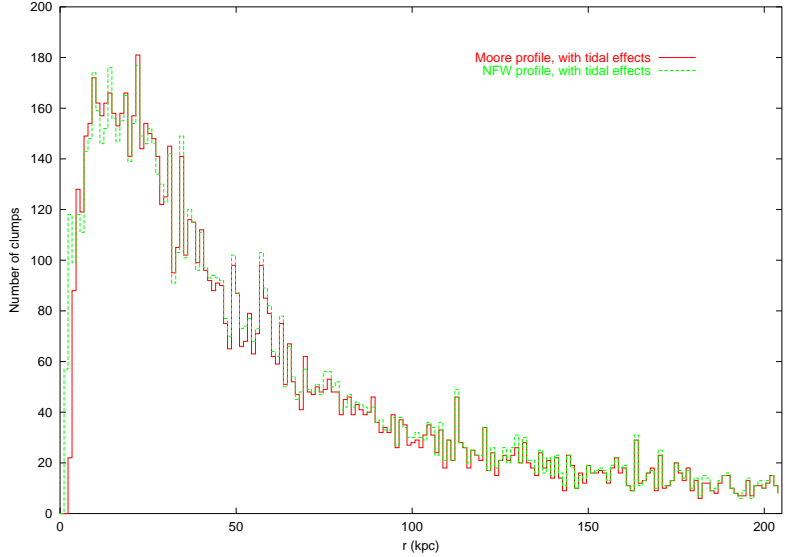}
\caption{\label{subnum}
Number of substructures along the radius of the MW for
the Moore and the NFW profiles.
}
\end{figure}

The tidal radius is defined as the radius of the substructure
at which the tidal forces of the host exceeds the self gravity of the
substructure. Assuming that both the host and the substructure gravitational
potential are given by point masses and considering the centrifugal 
force experienced by the substructure the tidal radius at the Jacobi limit
is given by \cite{hayashi}
\begin{equation}
r_{\text{tid}} = r_c \left( \frac{m}{3M_{vir}} \right)^{\frac{1}{3}}\ ,
\end{equation}
where $r_c$ is the distance of the substructure to the GC. 
The substructures with $r_{\text{tid}} \lesssim r_s$ will be disrupted
completely and be discarded in our realization of the substructure population.
The mass of a substructure is also recalculated by subtracting the mass
beyond the tidal radius. 
Therefore the final radial distribution of substructures near the GC is 
somewhat different
from that given in Eq. (\ref{prob}) which is shown in Fig. \ref{subnum}.
Indeed the substructures near the GC are disrupted completely after we take
the tidal effects into account. The 
substructures with NFW profile can exist more near the GC than the Moore
profile. This is because that the NFW profile is more centrally concentrated
with smaller $r_s$.

\subsection{Concentration parameter}

Once the population of the substructures are determined we need to
determine the dark matter density profile in each substructure
in order to calculate the annihilation of neutralinos.
The density profile in the substructure, 
either the NFW or the Moore profile, is characterized by two
scale parameters $\rho_s$ and $r_s$. The two parameters can be
determined using two different methods. In the first method they are
determined by one virial parameter, the virial mass (or equivalently
the virial radius, or virial velocity), and the concentration
parameter, which relates the virial and the inner scale radius. 

N-body simulation shows that the concentration of the substructure
is strongly correlated with the formation epoch of the substructure.
At an epoch of redshift $z_c$ a typical collapsing mass $M_{*}(z_c)$
is defined by $\sigma[M_{*}(z)]=\delta_{sc}(z)$, where the 
 $\sigma[M_{*}(z)]$ is the linear rms density fluctuation on the
comoving scale encompassing a mass $M_*$, $\delta_{sc}$ is the
critical overdensity for collapsing  at the spherical collapse model.
The collapsing mass  $M_*$ is determined by the primordial linear
power spectrum of the fluctuations and the known cosmology, which
determines the evolution of the fluctuations. In a semi-analytic model 
Bullock et al.\cite{bullock01} relate the typical collapsing mass to
a fixed fraction of the virial mass of a halo $M_{*}(z_c)=FM_{\text{vir}}$.
The concentration parameter of a halo with virial mass $M_{\text{vir}}$
at redshift $z$ is then determined as
$c_{\text{vir}}(M_{\text{vir}},z)=K\frac{1+z_c}{1+z}$. Both $F$ and $K$
are constants to fit the numerical simulations. 
Once $F$ and $K$ are determined the concentration of a halo is completely
determined by the cosmology in hand. We notice that a smaller $M_{\text{vir}}$
corresponds to a smaller collapsing mass and early collapsing epoch when
the Universe is denser and therefore a larger concentration parameter. 
Therefore smaller subhalos are more closely concentrated. 

\begin{figure}
\includegraphics[scale=0.7]{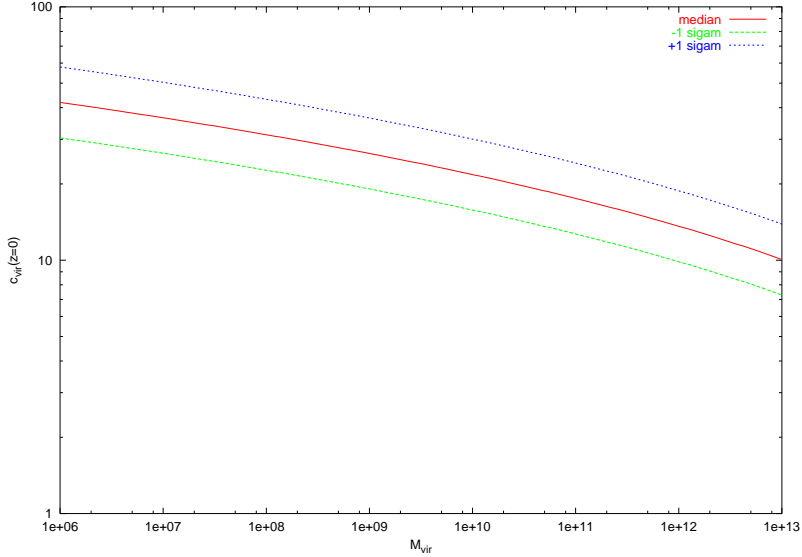}
\caption{\label{concen}
Concentration parameter as a function of the virial mass  of the halo
calculated according to the Bullock model\cite{bullock01}.
The model parameters are taken as $F=0.015$ and $K=4.4$. 
The cosmology parameters are taken as  $\Omega_M=0.3$, $\Omega_\Lambda=0.7$,
$\Omega_Bh^2=0.02$, $h=0.7$, $\sigma_8=0.9$ with three generations of 
massless neutrinos.
}
\end{figure}

In Fig. \ref{concen} 
we plot the concentration parameter at $z=0$ as a function of the 
virial mass of a halo according to the Bullock model\cite{bullock01}.
In the calculation, we have taken a standard scale invariant 
primordial spectrum of the fluctuation 
with the cosmology parameters taken as $\Omega_M=0.3$, $\Omega_\Lambda=0.7$, 
$\Omega_Bh^2=0.02$, $h=0.7$,  $\sigma_8=0.9$ and
three generations of massless neutrinos.
The model parameters are taken as $F=0.015$ and $K=4.4$.
The scatter of the concentration parameters for a given halo
mass is log-normal with $1\sigma$ deviation around the mean as $\Delta
(\log_{10}c_{\text{vir}})=0.14$. In Fig. \ref{concen} both the
median and the $\pm 1\sigma$ values of the concentration parameters
are plotted.
                                                                                
The Bullock model reproduces the concentration parameters quite well
from the N-body simulations. From Fig. \ref{concen} the experiential
formula is confirmed again that $c_{\text{vir}}
\propto M_{\text{vir}}^{-\beta}$. We expect that this exponential
relation of the concentration parameter 
and virial mass for subhalos should
be very well followed, since subhalo forms early at the epoch
when the Universe is dominated by matter with approximate 
power-law power spectrum of fluctuations.
Besides the Bullock model we have also 
adopted other recent simulation results in the literature. 
We use the experiential relation between the concentration 
parameters and virial mass and fit the parameter $\beta$ from 
these simulation results.
The  density profile of the substructure and furthermore the 
$\gamma$-ray flux from the substructure are then calculated.
We find the concentration parameter is the most sensitive parameter
in determining the $\gamma$-ray flux.
Different behavior of the concentration leads to different predictions 
of the $\gamma$-ray fluxes.

The second method to determine the profile parameters, 
as given in \cite{tasit}, requires each substructure produced
following Eq. (\ref{prob}) is compact
enough to resist the tidal stripping. 
Since the distribution of substructures
in Eq. (\ref{prob}) is given by simulations at $z=0$ 
the second method reflect the simulation results faithfully in the 
statistical meaning. Since the simulation
can not resolve the region near the GC we have given a cutoff at
$r=10$ kpc from the GC requiring no subhalo exits inside the cutoff radius. 
The conditions to determine the parameters are given by\cite{tasit}:
$R_{\text{vir}}=R_{\text{tid}}$, $\int_0^{R_{\text{vir}}}\rho(r)dV=M_{sub}$
and $\rho_{host}(r_{cl})=\rho_{cl}(R_{\text{vir}})$.
The last condition requires that the density of the subhalos at $R_{vir}$
should equal the local density of the host halo at the position of the
subhalo $r_{cl}$.

\subsection{Results}

\begin{figure}
\includegraphics[scale=0.7]{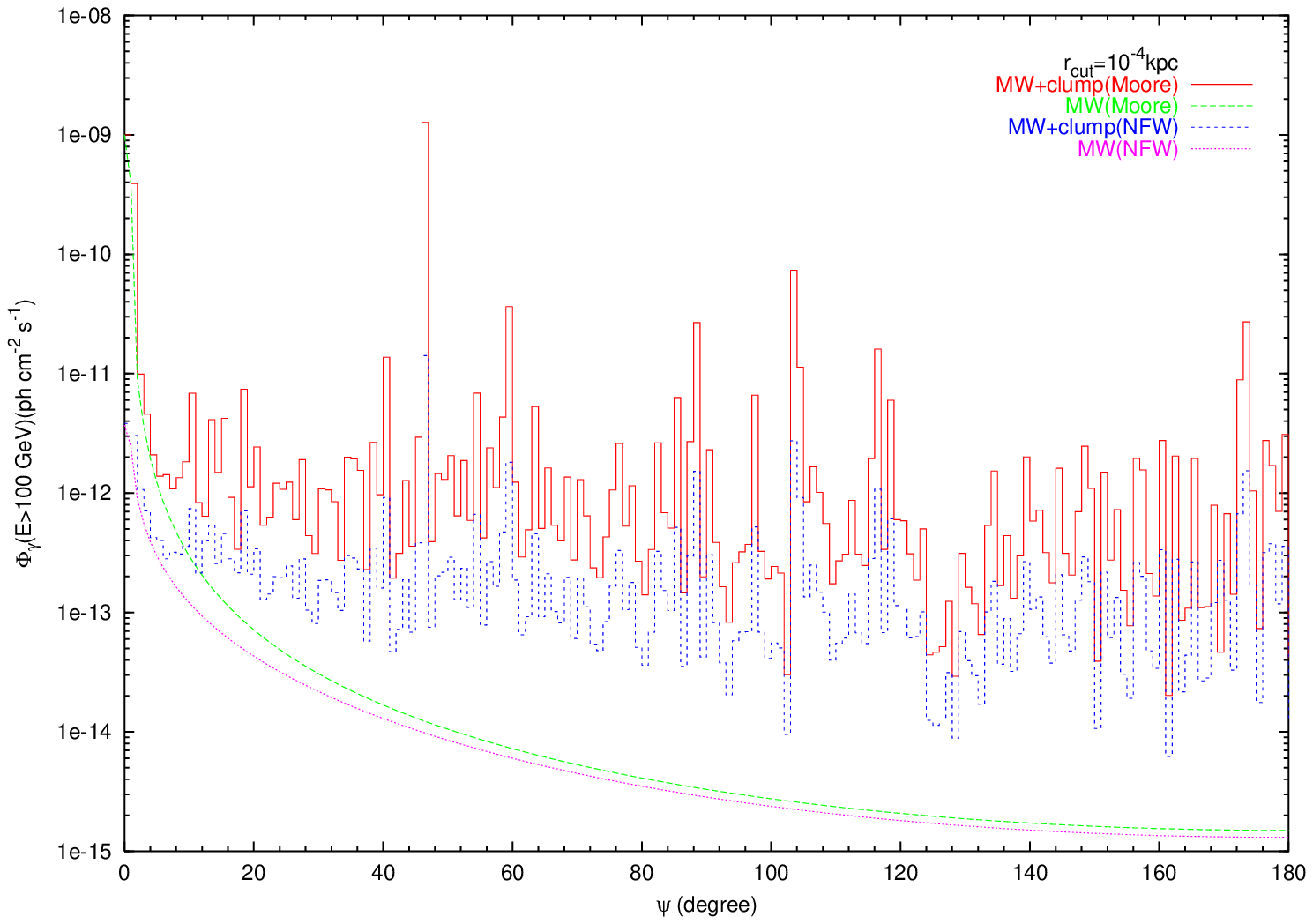}
\includegraphics[scale=0.7]{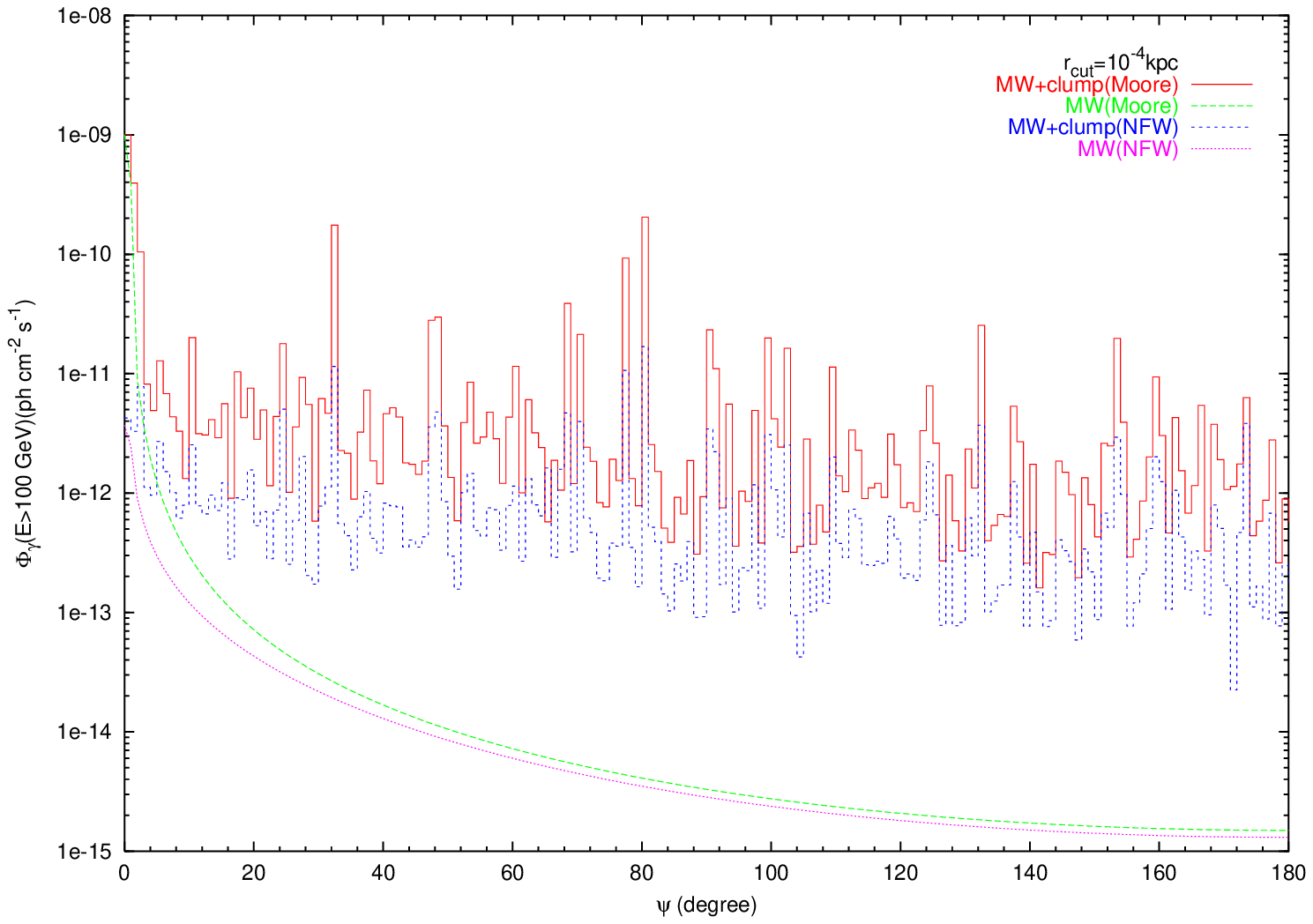}
\caption{\label{one}
The integrated $\gamma$-ray fluxes above $100 GeV$ as function of
the angles of the sources relative to the GC. In the upper panel we adopt 
the concentration parameters of the subhalos given in \cite{bullock01}
while in the lower panel we adopt that in \cite{reed}.
Both contributions from the smooth
component of DM and from subhalos are plotted. 
}
\end{figure}

We can now present the results of the $\gamma$-ray
fluxes from the neutralino annihilation in the MW substructures. 
The fluxes are averaged within the angular resolution of $1^\circ$
in the following way: if a subhalo is too small or too far from the
detector that its angular size is smaller than the angular resolution
we calculate the annihilation flux of the whole subhalo; 
on the contrary for these subhalos the detector can resolve 
we calculated the flux within 
the $1^\circ$ angular resolution assuming that we are aiming at the
center of the subhalo. 
The annihilation core 
radius in the following calculations is taken as $r_{cut}=10^{-4} kpc$,
which is quite a conservative value and affects the the results
little according to Fig. \ref{core_radius}. 

In Fig. \ref{one} we  plot the integrated $\gamma$-ray
fluxes above the threshold energy of 100 GeV as a function of $\psi$, 
the angle of the source to the direction of the GC. In the figure
no information is shown about the other
direction around the axis along the Sun and the GC. 
Only the maximal $\gamma$-ray flux at this direction is chosen 
for each $\psi$ and plotted in the figure.
Both the $\gamma$-rays annihilated from the smooth dark halo and that by
adding the emission from the subhalos and the smooth component 
are shown for the Moore and the NFW profiles. 
For the smooth component
the $\gamma$-ray flux decreases rapidly when departing from the GC while
for the subhalos the contribution seems almost isotropic to different
directions. This is due to the fact that the Sun is only $8.5$ kpc away
from the GC. In the upper
panel we calculate the flux adopting the Bullock model\cite{bullock01} and 
in the lower panel we adopt the concentration parameter according to the
simulation by Reed et al. \cite{reed}. 
Generally the Moore profile predicts greater $\gamma$-ray fluxes.
At the GC the $\gamma$-ray flux for the Moore profile is about 300
times higher than that for the NFW profile.  However, 
for the subhalos the difference between the
two profiles is only about one order of magnitude. Therefore the 
uncertainty of the predicted $\gamma$-ray fluxes from the subhalos
due to the cosmological factor is much smaller than that from the GC. 
The figure can be used to other threshold energies by
a global shift of the particle factor from the Fig. \ref{GC_flux}.

\begin{figure}
\includegraphics[scale=0.7]{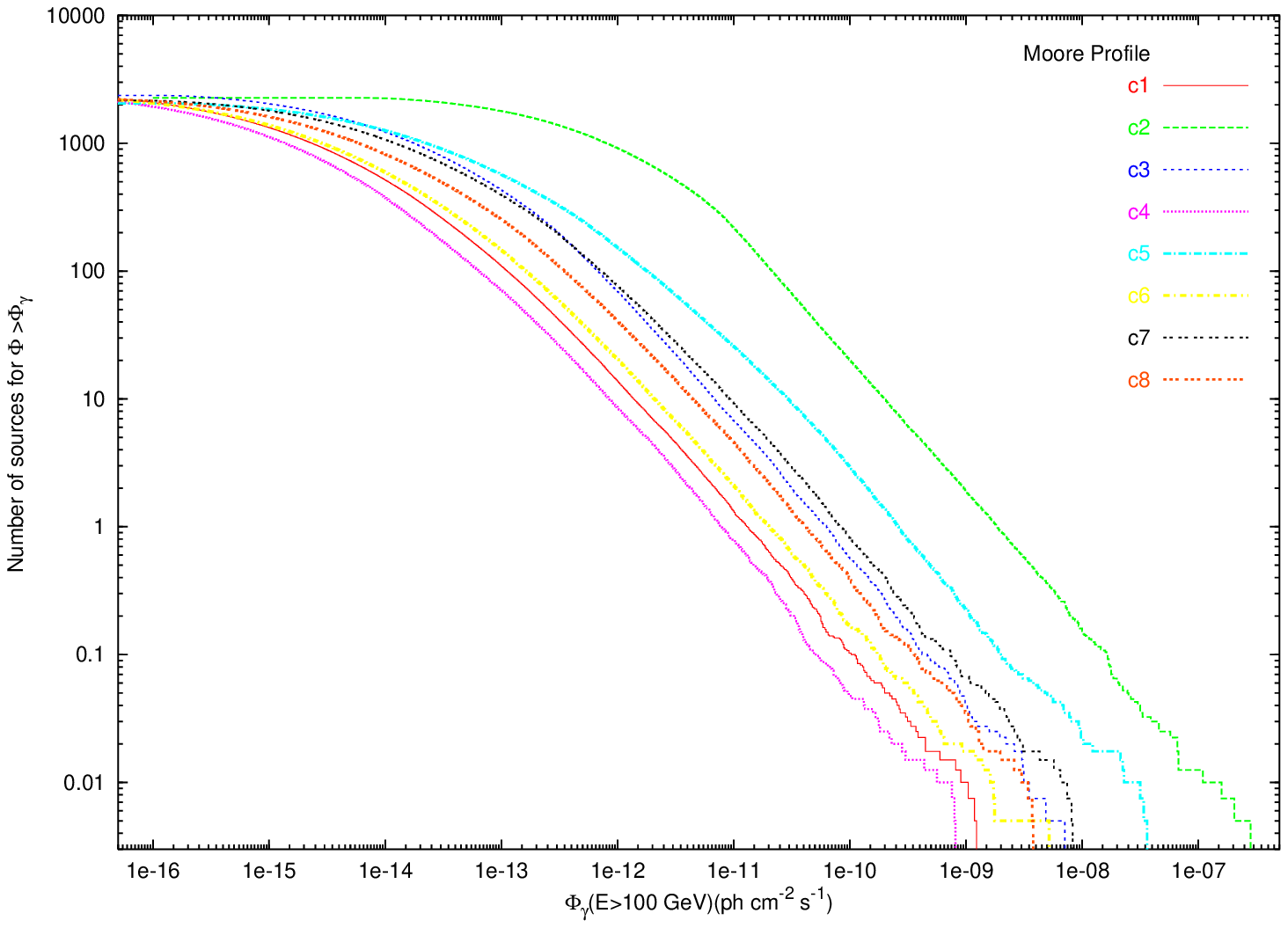}
\includegraphics[scale=0.7]{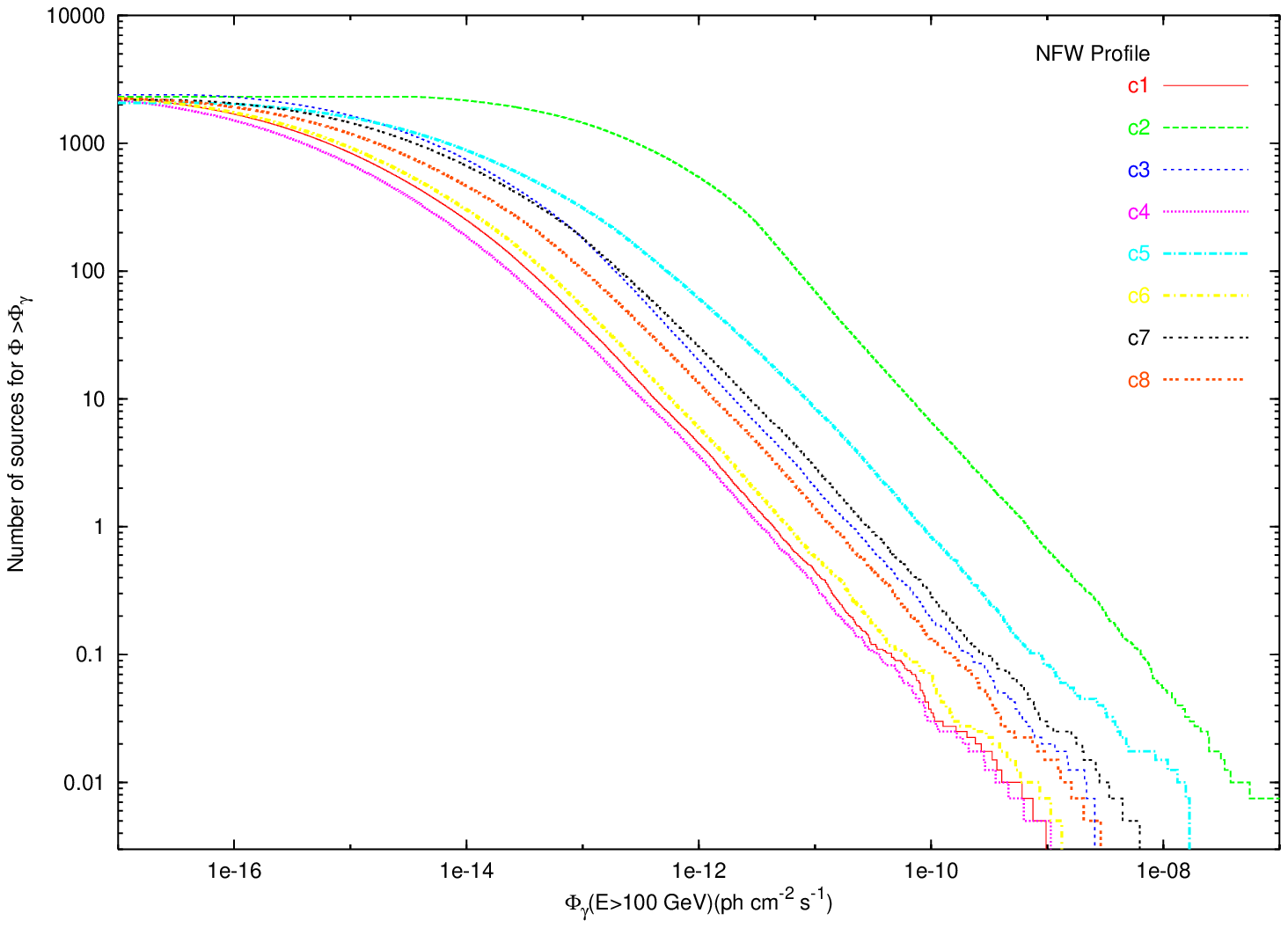}
\caption{\label{result}
The cumulative number of subhalos as function of the
integrated $\gamma$-ray fluxes $n(>\Phi_\gamma)$ for the Moore
profile (upper panel) and the NFW profile (lower panel). 
The results are given within the zenith angle of $60^\circ$.
The curves represent the results according to 
different simulations as explained in the text.
These curves give the number of subhalos which emit $\gamma$-rays
with the integrated flux above $\Phi_\gamma$.
}
\end{figure}
                                                                                
It should be noted that Fig. \ref{one} illustrates 
the $\gamma$-ray fluxes from the subhalos due to
a statistical realization of our Galaxy. None of the
positions of peaks are predicted exactly and the maximal flux 
may have a large
fluctuation: it is possible that a subhalo is accidentally
located near the solar
system. Therefore we try to calculate the statistically averaged 
fluxes by realizing one hundred MW sized halos and count the
average number of subhalos emitting $\gamma$-rays above any
intensities. 

Fig. \ref{result} gives the cumulative number of subhalos emitting
$\gamma$-rays with intensity above the integrated flux $\Phi_\gamma$. 
It should be noted that the results, also for other figures hereafter, 
are given within the zenith angle of
$60^\circ$, which is the maximal angle a ground array can possibly probe, 
instead of the whole sky. In the upper
panel we plot the results for the Moore profile while the lower
panel is for the NFW profile. The curves are given by calculating
the density profile of subhalos according to different
author's simulation results, where `c1' denotes the simulation of 
Ref. \cite{klypin};
`c3' of Ref. \cite{reed}; `c4' of Ref. \cite{eke} for the
$\Lambda$CDM model with $\sigma_8=0.9$;
`c6' uses the median $c_{vir}-M_{vir}$ relation for distinct halos of the
Bullock model given in Ref. \cite{bullock01}, while 
`c7' and `c8' take the upper $2\sigma$ and $1\sigma$ values of the same model
respectively; `c2' adopts the simulation results 
given in dense matter environment
of the same reference and extends the relation to small subhalos; since this
relation gives very large concentration parameters for small subhalos we have
cut the concentration parameter arbitrarily if $c> 100$ in `c5'.
From this figure  we can easily read the number
of the expected detectable subhalos if the sensitivity of a 
detector is given (with same threshold energy and angular resolution
adopted here).

\begin{figure}
\includegraphics[scale=0.7]{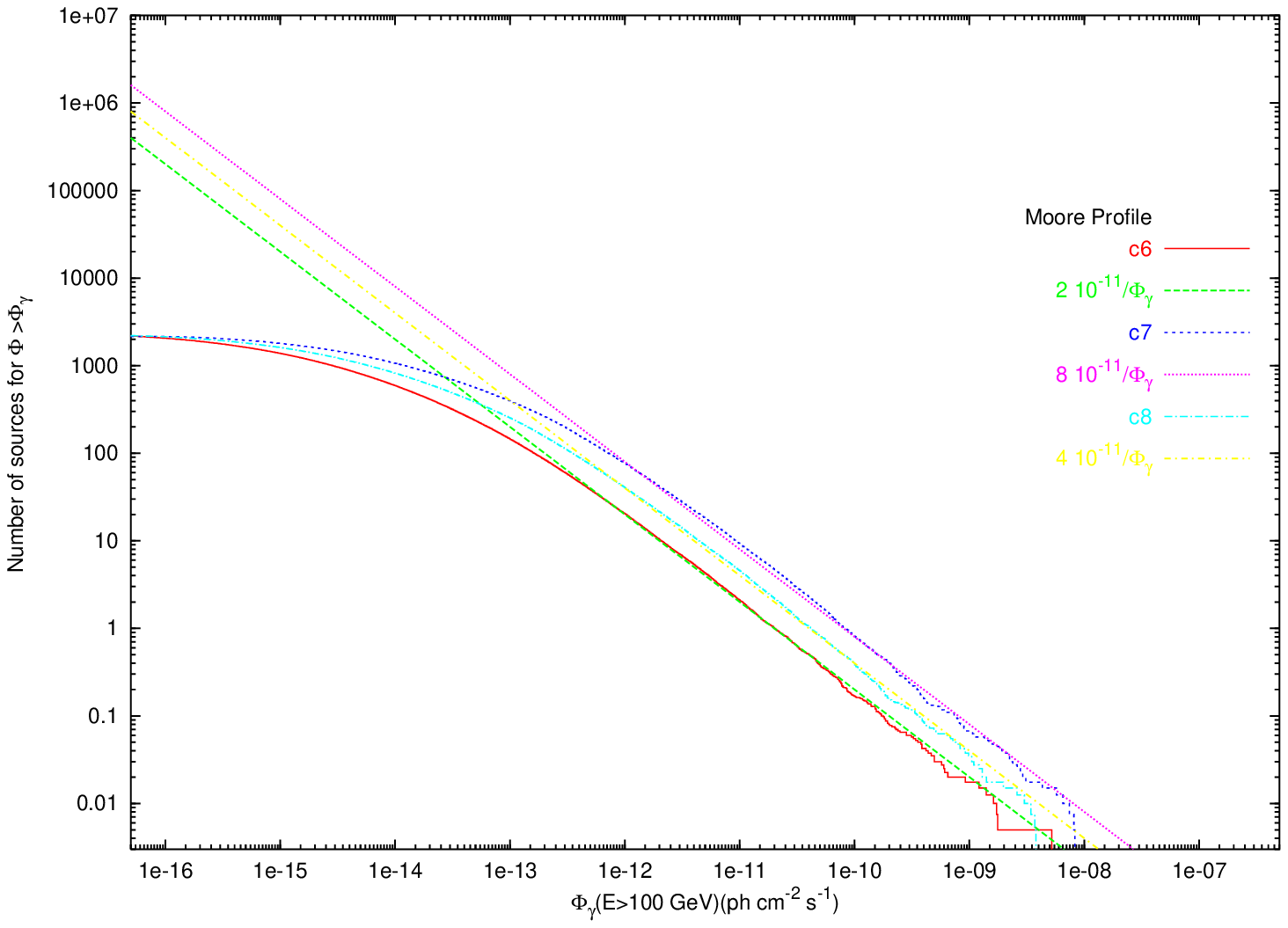}
\includegraphics[scale=0.7]{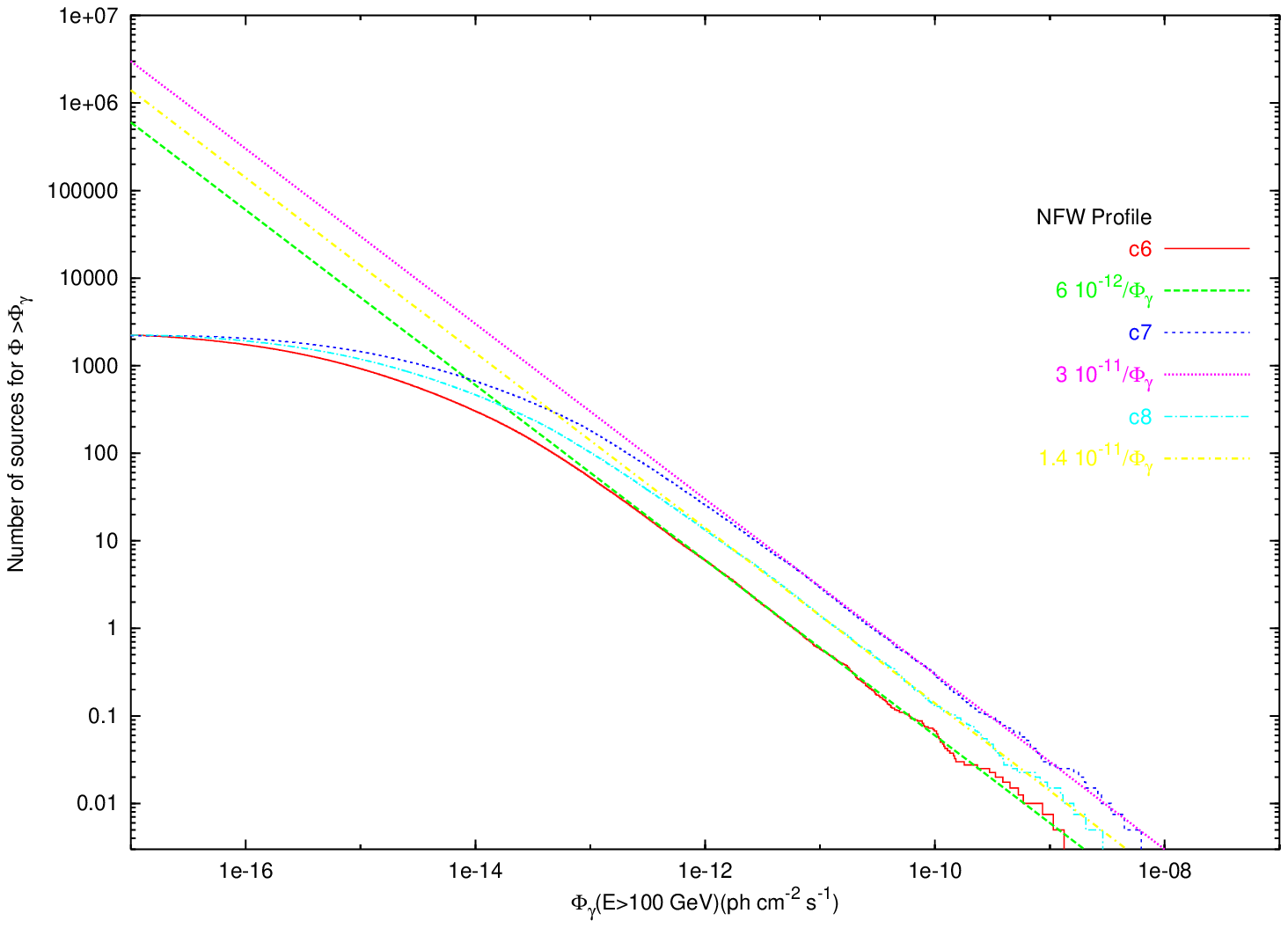}
\caption{\label{fit}
The cumulative number of subhalos as function of the
integrated $\gamma$-ray fluxes $n(>\Phi_\gamma)$ for the Moore
profile (upper panel) and the NFW profile (lower panel).
The results are given within the zenith angle of $60^\circ$.
The lines are the  corresponding fits of the curves for high
$\gamma$-ray fluxes. From these fits we can easily read out the
number of subhalos which emit $\gamma$-rays
with the integrated flux above $\Phi_\gamma$.
}
\end{figure}
                                                                                
\begin{figure}
\includegraphics[scale=0.7]{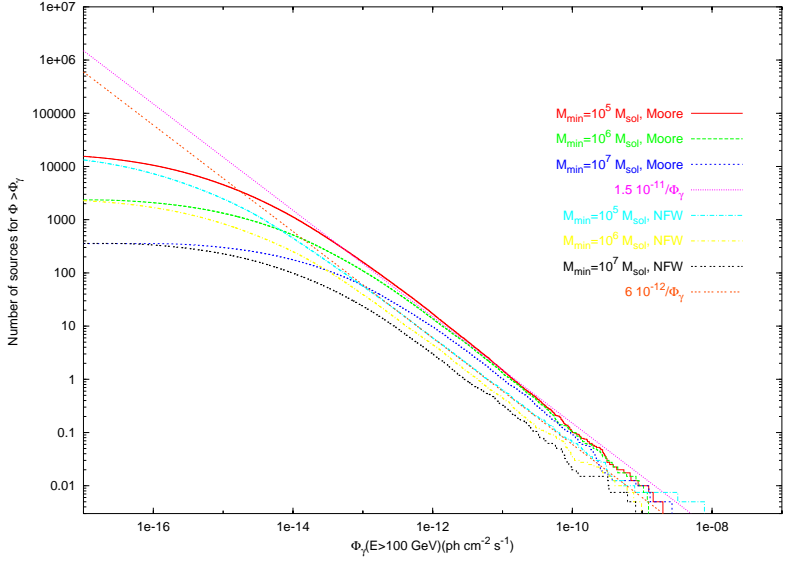}
\caption{\label{lowmass}
The cumulative number of subhalos as function of the
integrated $\gamma$-ray fluxes $n(>\Phi_\gamma)$ for
taking different cuts of the minimal mass of the subhalos.
The results are given within the zenith angle of $60^\circ$.
The lines are the  corresponding fits of the curves for high
$\gamma$-ray fluxes.
The results are quite insensitive to the cut of the
minimal subhalo mass.
}
\end{figure}

\begin{table}
\begin{tabular}{l| c c c c c c c c } \hline \hline
& c1  & c2   & c3 & c4  & c5 & c6  & c7  & c8 \\
\hline
 Moore & $1.5\times 10^{-11}$  & $2\times 10^{-9}$ & $6\times 10^{-11}$  & 
$8\times 10^{-12}$ & $3\times 10^{-10}$ &  $2\times 10^{-11}$  
&  $8\times 10^{-11}$  &  $4\times 10^{-11}$ \\
 NFW   &  $4.5\times 10^{-12}$  &  $7\times 10^{-10}$ & $2\times 10^{-11}$  &  $3.5\times 10^{-12}$  &  $9\times 10^{-11}$  & $6\times 10^{-12}$ &  $3\times 10^{-11}$  &  $1.5\times 10^{-11}$ \\
\hline \hline
\end{tabular}
\caption{
\label{fitn0}
Values of the constant $n_0$ in the fit 
$n(>\Phi_\gamma)=n_0/\Phi_\gamma$ to
the curves in Fig. \ref{result} for the Moore and the NFW profiles.
}
\end{table}

We find the cumulative number of subhalos for large fluxes 
can be well fitted by an inverse power law, 
$n(>\Phi_\gamma)=n_0/\Phi_\gamma$, as shown in Fig. \ref{fit}.
The deviation of the calculated curves from the
fitted lines at the end of largest fluxes is due
to the fact that we do not have enough statistics here, while the deviation
at the end of lowest fluxes may be due to the cut of the minimal mass of 
subhalos. In the table \ref{fitn0} we give the fitted constant $n_0$ for 
different curves. The Moore profile predicts about 3 times more detectable
subhalos than the NFW profile averagely. 
Therefore the $\gamma$-ray fluxes from the subhalos are not so 
sensitive to the dark matter profiles as these from the GC, 
as shown in the Fig. \ref{one}.
In Fig. \ref{lowmass} we show how the cumulative number of subhalos
changes as the value we take for the minimal mass of subhalo. We
illustrate the result in the model `c1'. 
The dependence on the
minimal mass of subhalo is very weak, which can be understood from some
simple scaling arguments\cite{kou}. It is worth noting that 
taking smaller minimal subhalo mass indeed makes the curves closer to the
fitted lines at the end of smallest fluxes.

\begin{figure}
\includegraphics[scale=0.7]{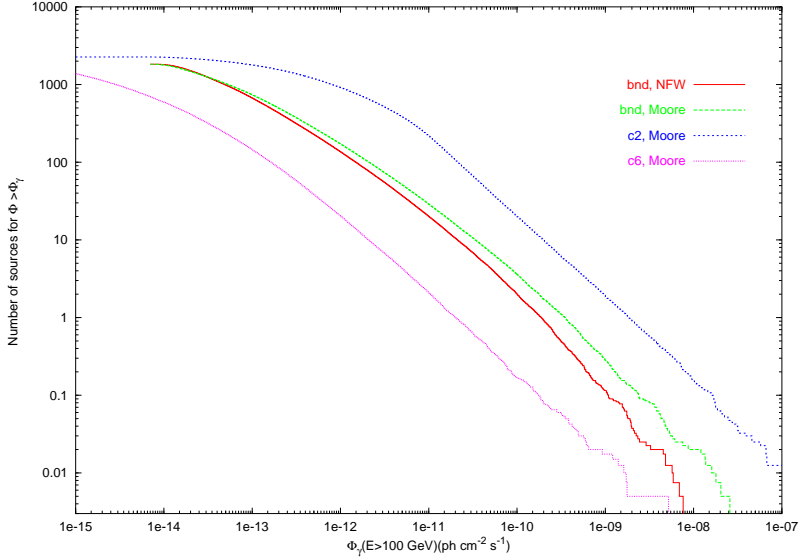}
\caption{\label{bound}
The cumulative number of subhalos as function of the
integrated $\gamma$-ray fluxes $n(>\Phi_\gamma)$.
The results are given within the zenith angle of $60^\circ$.
The curves are calculated according to the second method
to determine the profile parameters as described in the text.
The corresponding results from the models `c2' and `c6' are
also shown.
}
\end{figure}

Finally, in Fig. \ref{bound} we give the results according to the second
method of determining the concentration parameter. Both the NFW and Moore
profiles predicts much larger fluxes compared with that from the model
`c6' using the Moore profile, while smaller than that of the `c2' model.

In summary,
we calculated the $\gamma$-ray fluxes from 
the MW substructures in this section. There are extensive studies
on the evolution and distribution of substructures by numerical
simulations. We integrate the recent simulation results in our
calculation. The different simulation 
results give uncertainties in predicting
the $\gamma$-ray fluxes. The concentration parameter is the most
sensitive parameter in determining the $\gamma$-ray fluxes.
We give the statistical average number of subhalos with $\gamma$-ray
intensity above some values. The result is well fitted by an inverse
power law. From these results the number of detectable subhalos
is easily read out if the sensitivity of a detector is known.
We will discuss the detectability of the $\gamma$-ray signals
from the MW subhalos in the next section.

\section{ detectability }

The detectability of a signal is defined by the ratio of 
the signal events to the fluctuation of the background.
Since the background follows the Poisson statistics, its
fluctuation has the amplitude proportional to $\sqrt{N_B}$.
The \textit{significance} of the detection is quantified by
$\sigma=\frac{n_\gamma}{\sqrt{N_B}}$. 

The signal events are given by
\begin{equation}
\label{ngamma}
n_\gamma= \epsilon_{\Delta \Omega}\int_{E_{th},\Delta \Omega} A_{eff}(E,\theta)
\phi(E) dE d\Omega dT\ \ ,
\end{equation}
where $\epsilon_{\Delta \Omega}=0.68$ is the fraction of signal
events within the angular resolution of the instrument and the
integration is for the energies above the threshold energy $E_{th}$, within
the angular resolution of the instrument $\Delta \Omega$ and for the
observational time. The effective
area $A_{eff}$ is a function of energy and zenith angle.
The $\phi(E)$ is the flux of $\gamma$-rays from DM annihilation.

The corresponding 
expression for the background is similar to Eq. (\ref{ngamma}).
The background includes contributions from the hadronic and electronic
comic-rays and the Galactic and extragalactic $\gamma$-ray emission.
We have adopted the expressions as
\begin{equation}
\phi_h(E) =1.49 E^{-2.74} cm^{-2}s^{-1}sr^{-1}GeV^{-1}
\end{equation}
for the hadronic contribution \cite{gaisser},
\begin{equation}
\phi_e(E) =6.9\times 10^{-2} E^{-3.3} cm^{-2}s^{-1}sr^{-1}GeV^{-1}
\end{equation}
for the electronic contribution \cite{longair},
\begin{equation}
\label{extra}
\phi_{\text{extra}-\gamma}(E) =1.38 \times 10^{-6}E^{-2.1} cm^{-2}s^{-1}sr^{-1}GeV^{-1}
\end{equation}
for the extragalactic $\gamma$-ray emission extrapolated from EGRET
data at low energies\cite{extra} and 
\begin{equation}
\phi_{\text{galac}-\gamma}(E) =N_0(l,b)\times 10^{-6} E^{-2.7} cm^{-2}s^{-1}sr^{-1}GeV^{-1}
\end{equation}
for the Galactic $\gamma$-ray emission, also extrapolated from the EGRET
data at low energies\cite{gal},
with $N_0$ the normalization factor depending on galactic coordinates $(l,b)$.
The $N_0$ is modeled using EGRET data at 1 GeV \cite{gal}
\begin{equation}
N_0(l,b)=\frac{85.5}{\sqrt{1+\left(l/35\right)^2}\;\sqrt{1+\left(b/(1.1+|l|\,0.022)\right)^2}}
 \,+\,0.5
\end{equation}
for $|l|\, \geq 30^{\circ}$ and
\begin{equation}
N_0(l,b) =\frac{85.5}{\sqrt{1+\left(l/35\right)^2}\;\sqrt{1+\left(b/1.8\right)^2}}\,+\,0.5
\label{n0}
\end{equation}
for $|l|\,\leq 30^{\circ}$, with the longitude $l$ and the latitude
$b$ varying in the intervals $[-180^{\circ}, 180^{\circ}]$ and 
$[-90^{\circ}, 90^{\circ}]$, respectively.

Since most background comes from the hadronic cosmic rays, 
the hadron-photon identification efficiency is an important factor to
reduce the physical background. For a satellite borne experiments,
such as 
GLAST \cite{glast},
an identification efficiency of charged particles as high as $99.997\%$
can be assumed, while $90\%$ for the photons\cite{geff}. However, the
effective area of this kind of experiments is limited by the size 
of the satellite and has the order of $A_{eff}\sim 1 m^2$.
The atmospheric \v{C}erenkov telescopes (ACT), such as VERITAS\cite{veritas}, 
MAGIC\cite{magic} and HESS\cite{hessh}, can have very large effective
area with an identification efficiency of $99\%$ for both the 
hadronic and the electromagnetic primary particles. However, the ACTs have 
a small field of view ($\sim 5^\circ$) with a duty cycle of about 10\%.
Therefore the ACTs are suitable for the observation of point sources and
can not do the blind search. 

The ground-based extensive air shower (EAS) arrays, such as ARGO\cite{argo},
MILAGRO\cite{milagro} and the next generation all-sky high 
energy $\gamma$-ray telescope HAWC\cite{hawc}, have complementary
properties to the satellite borne experiments and the ACTs. They also
have large effective areas and at the same time they have large
filed of view ($\sim 60^\circ$) and a duty cycle of about 100\%.
However, the EAS arrays have low hadron-photon identification efficiency.
For the ARGO we assume no discrimination between the hadron
and the photon, while for the HAWC the hadron-photon discrimination
can improve the significance of the detection by a quality factor of 
1.6 \cite{hawc}.

\begin{figure}
\includegraphics[scale=0.8]{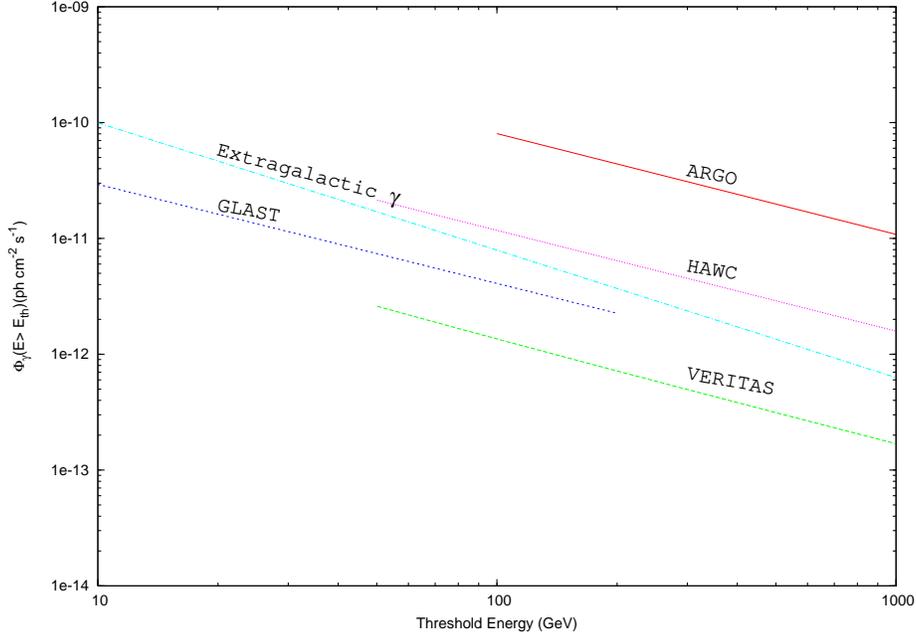}
\caption{\label{sensitive}
The $3\sigma$ sensitivity curves of different 
detectors as functions of the threshold
energy of the $\gamma$-rays. The curves are calculated according to
the parameters given in the text.}
\end{figure}
                                                                                
In Fig. \ref{sensitive} we show the $3\sigma$ sensitivity curves of 
different detectors as function of the threshold energy. 
We have assumed the angular resolution
as $0.1^\circ$ for GLAST and VERITAS 
and the angular resolution as $1^\circ$ for ARGO
and HAWC. For GLAST, ARGO and HAWC we use
1 year of data taking, while  for VERITAS 250 hours observation 
pointing to a source is assumed. 
We make rough approximation of energy independent effective area,
$10^3 m^2$, $3\times 10^4 m^2$ and $4\times 10^4 m^2$ for 
ARGO, HAWC and VERITAS respectively. The diffuse extragalactic $\gamma$-ray
background from Eq. (\ref{extra}) is also plotted in the figure
for  comparison. 

The detectability of GLAST and VERITAS for the $\gamma$-rays
from the Galactic subhalos
has been carefully studied in the Ref. \cite{kou,peirani}. It is shown that
the significance of the detection of the annihilated $\gamma$-rays
at GLAST is maximized for $M_\chi\approx 40 GeV$, since GLAST has
low energy threshold and thus introduces large background and 
the $\gamma$-ray
flux decreases when increasing the neutralino mass. For neutralino
heavier than  $100 GeV$ GLAST has little chance to detect the signal\cite{kou}.
We can get similar conclusion from Fig. \ref{GC_flux} that for small
threshold energy the maximal flux is at the lower end of the neutralino
mass while as the threshold energy becomes higher the position of the
maximal flux moves to a neutralino mass of several hundred GeV.
VERITAS, with
a higher threshold energy, is sensitive enough to detect 
the $\gamma$-rays annihilated from heavy neutralinos. However, its
small field of view limits its ability to do such observation\cite{kou}.

However, it is possible to do the observation by the EAS array according
to our calculations. Taking the threshold energy as $100GeV$,
ARGO and HAWC require the $\gamma$-ray fluxes as high as 
$8\times 10^{-11}cm^{-2}s^{-1}$ and $1\times 10^{-11}cm^{-2}s^{-1}$,
respectively, to achieve a $3\sigma$ significance detection.
Comparing with the Table I, we get that ARGO can detect 0.1 to as many as
about 25 subhalos with a Moore profile and from 0.04 to about 9 subhalos
with a NFW profile for one year data taking. For HAWC, since its sensitivity
is about 8 times higher than that of ARGO it can also detect
more subhalos by 8 times than that of ARGO. 
For example, 
for the Bullock model and its $2\sigma$ upper limit, ARGO has the
ability to detect $1/4$ and $1$ subhalos respectively for the Moore profile
and about $0.1$ to $0.4$ subhalos respectively for the NFW profile. While
for HAWC even with NFW profile and the median concentration parameter of
the Bullock model it can detect about $1$ subhalo for one year observation.

Here we can conclude that for the neutralino dark matter as light
as about $100 GeV$ 
GLAST is most suitable to observe the annihilated $\gamma$-rays
from the Galactic subhalos. For the heavier neutralino dark matter 
$\sim 500 GeV$ ($< 1 TeV$) ARGO/HAWC can be a viable 
complementary. Especially, if HAWC is built it has great potential
to do the observation.

\section{discussion and conclusions}

In this work we calculated the $\gamma$-ray fluxes produced by
the dark matter annihilation in the Galactic subhalos and
discussed the detectability of such signals by different types
of detectors according the nature of the dark matter. 

We explored the low energy parameter space of the MSSM and studied
the uncertainties from the particle physics in predicting the
$\gamma$-ray fluxes. Uncertainties from the astrophysics are 
also carefully studied, where 
we find the most sensitive parameter in predicting the
$\gamma$-ray fluxes is the concentration parameter of subhalos.
At the moment there exist discrepancies according
to different author's numerical simulations. We present the results
according to several recent simulation results.
 
Assuming optimistic SUSY parameters the $\gamma$-rays from 
subhalos may be detected by 
satellite borne experiment, such as GLAST, which has large field of
view and small effective area if the $\gamma$-ray flux is large enough
\cite{kou} (when the neutralino is light $\lesssim 100 GeV$).  
On the contrary,
when the neutralino mass is large ($\sim 500 GeV$) the $\gamma$-ray flux
is reduced and only ground based experiments with large effective
area and large field of view, such as ARGO/HAWC, 
can do the job. We calculated the statistic average numbers
of the detectable subhalos at these detectors. 

If such an annihilation signal is indeed detected in the future it will not
only indicate the weakly interaction between the dark matter
particles and further implicate the nature of new physics beyond the SM 
but also tell us 
a lot about the nature of the subhalos: they must have a cuspy profile with
the Moore or the NFW form or some form between them and the CDM scenario is
favored without power suppression at the subgalactic scale. However,
from this single measurement
neither the SUSY parameter space nor the subhalos distribution and
its profile can be actually determined, since
both sides still have large uncertainties. 
Anyway, the indirect search of dark matter provides valuable 
complementary to both the collider study of particle physics and
the more precise simulations of the dark matter evolution.


Finally we want to stress again the advantages of search for 
the dark matter annihilation from the MW subhalos. 
First, subhalos produce clean annihilation signals, as we have
explained before. The annihilation radiation from the
GC is heavily contaminated by the baryonic processes. Furthermore, 
the density profile near the GC is  complicated due
to the existence of baryonic matter. For example, the SMBH can
either steepen or flatten the slope of the DM profile at the innermost
center of the halo\cite{ullio}. For subhalos, their profile may
simply follow the simulation results. 
Second, the small
subhalos form earlier and have larger concentration parameter, which
leads to relatively greater annihilation fluxes. 
Third, the DM profile may be not universal, as shown in the simulation
given in Ref. \cite{reed,jing}. Smaller subhalos have steeper central cusp.
In this case, from Fig. \ref{one} taking the GC the NFW profile and the
subhalos the Moore profile, the $\gamma$-ray fluxes from the
subhalos may even be greater than that from the GC.
Forth, according to the hierarchical formation of structures 
in the CDM scenario we expect that subhalos should contain their
own smaller sub-subhalos, which can further enhance the annihilation
flux. The sub-subhalos have been observed in the numerical simulations,
such as in the Ref. \cite{zentner05}.
Finally, the environmental trend seems to make the subhalos more concentrated 
\cite{bullock01}. However, the effects need further studies by more precise
simulations.

\begin{acknowledgments}
We thank H.B. Hu, Y.P. Jing, L. Gao, X.M. Zhang, H.S. Zhao and X.L. Chen 
for helpful discussions. 
This work is supported by the NSF of China under the grant No.
10575111, 10120130794, 10105004.
\end{acknowledgments}

\end{document}